\documentclass{ws-rv10x7}
\usepackage{ws-rv-van}     % numbered citation/references (default)
\usepackage{ws-rv-thm}     % comment this line when `amsthm / theorem / ntheorem` package is used
\usepackage{subfigure}
\usepackage{bm}
\usepackage{comment}
\makeindex
\begin{document}

\renewcommand{\thechapter}{23}

\chapter[23. Replica symmetry breaking in random lasers]{Replica symmetry breaking in random lasers: \\ experimental measurement of the  overlap distribution\label{ch23}}

\author[C. Conti, N. Ghofaniha, L. Leuzzi, G. Ruocco]{Claudio Conti$^{1,2}$, Neda Ghofraniha$^{2}$, Luca Leuzzi$^{1,3}$\footnote{luca.leuzzi@cnr.it}, Giancarlo Ruocco$^{1,4}$}

\address{1. Physics Department Sapienza University of Rome, Piazzale Aldo Moro 5 - 00185, Rome, Italy.}
\address{2. Institute of Complex Systems, National Research Council of Italy (CNR-ISC), Piazzale Aldo Moro 5 - 00185, Rome, Italy.}
\address{3. Institute of Nanotechnology, National Research Council of Italy (CNR-NANOTEC), Soft and Living matter Lab, Piazzale Aldo Moro 5, 00185, Rome, Italy.}
\address{4. Center for Life Nano- \& Neuro-Science, Italian Institute of Technology (CLN2S@Sapienza, IIT), Viale Regina Elena, 291 - 00161 Rome, Italy}

\begin{abstract}
In this chapter we report on the measurements of the overlap distribution of the replica symmetry breaking solution in complex disordered systems. After a general introduction to the problem of the experimental validation of the Parisi order parameter, we focus on the systems where the measurement has been possible for the first time: random lasers. Starting from first principles of light-matter interaction we sketch the main steps leading to the construction of the statistical mechanical model for the dynamics of light modes in a random laser, a spherical multi-p-spin model with complex spins.
A new overlap is introduced, the intensity fluctuation overlap, whose probability distribution, under specific assumptions, is equivalent to the Parisi overlap distribution. The experimental protocol for measuring this overlap is based on the possibility of experimentally realizing real replicas. After a description of the first experiment on the random laser made of  T5CO$_x$ grains we review and discuss various experiments measuring the overlap distribution, as well the possible connection with Levy-like distribution of the intensity of the light modes around the laser threshold, the connection with turbulence in fiber lasers and the role of spatial etherogeneities of light modes in random media. 
\end{abstract}

\body

\section{Introduction}\label{sec1}

\begin{figure}[t!]
\begin{center}
\includegraphics[width=.9\textwidth]{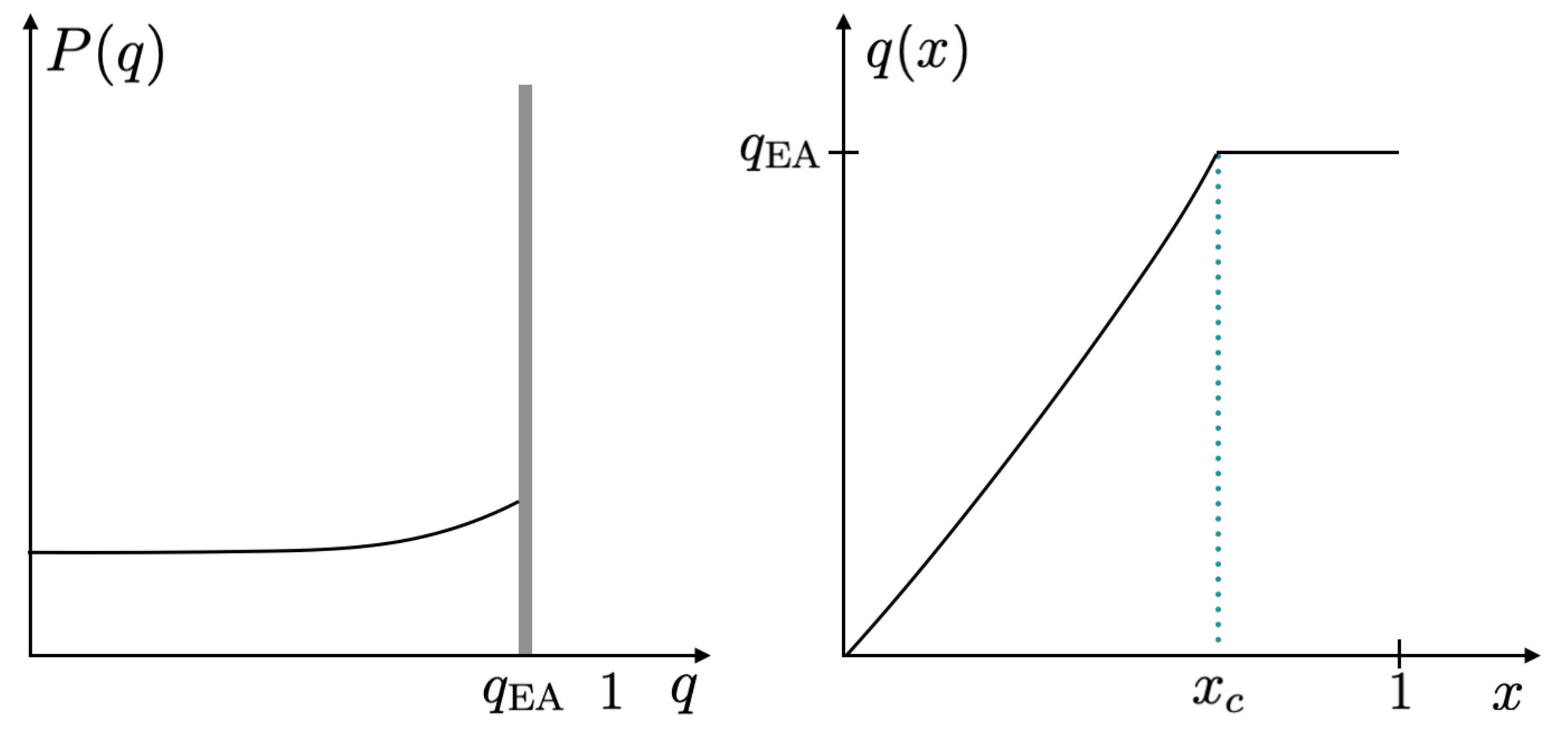}
\end{center}
 \caption{Left: sketch of a full replica symmetry breaking probability distribution of overlaps between states. The Edwards-Anderson parameter $q_{\rm EA}$ denotes the modal of the distribution, its meaning being the overlap of a state with itself. The thick grey bar represents a Dirac delta on $q=q_{\rm EA}$. 
 Right: the functional order parameter $q(x)$, i.e., the inverse of the cumulative distribution (\ref{x_q}), for a full RSB system. The point $x_c$ of discontinuous derivative is Eq. (\ref{x_q}) computed at $q=q_{\rm EA}^-$.
}
\label{fig-Pq-sketch}
\end{figure}

The  theory of replica symmetry breaking (RSB) relies on an order parameter which is not a number or a vector, rather it is a function of a continous variable.\cite{Parisi80a,Parisi80b} This is the main novelty of the theory of Parisi: to conceive and construct  an order parameter able to identify a thermodynamic phase of a system with multi-equilibria, a signature for complex disordered systems. The function in question is the probability distribution $P(q)$ of the overlap $q$ between equilibrium states, or its cumulative
\begin{equation}
x(q) = \int^q dq'\ P(q'),
\label{x_q}
\end{equation} or, equivalently, the inverse of the cumulative $q(x)$, see Fig. \ref{fig-Pq-sketch}. 
Numerous and astonishing are the outcomes  of this idea, right in the original replica formalism, as well as in further reformulations, such as the cavity method \cite{Mezard86, Mezard86b, Mezard01}.
Though $40$ years have passed since the conception of such quantity in spin-glasses, to experimentally observe  the order parameter in its full functional glory is still a rather difficult and challenging task.

 %--------------Figure ------------------------%

%----------------end figure -----------------%

In this paper, after exploring the early attempts to experimentally expose the inner structure of the organization of states predicted by the RSB theory in complex disordered systems, such as spin-glasses and structural glasses, we will show how, with some preliminar theoretical work, the distribution of the values of the overlaps between states can be sampled in random lasers \cite{Wiersma95, Cao99, Cao03r, Wiersma08,Andreasen11, Viola18,Gomes21}.

A first measurement of a peculiar behavior compatible with the - soon incoming - Parisi picture was performed by Nagata, Keesom and Harrison in 1978\cite{Nagata79} on a spin glass alloy, the cuprate-manganese, CuMn.
There, the magnetic susceptibility was carefully measured in static magnetic fields using two procedures.
In the first one the system is cooled down  at nearly zero magnetic field and, then, at different temperatures the system is perturbed by a magnetic field and the response is acquired.
In the second protocol the CuMn is cooled down embedded in a constant uniform magnetic field. Then, at each temperature  the field is perturbed and the field cooled susceptibility   measured. 
Both the zero-field-cooled (ZFC) and the field-cooled (FC) susceptibility behaviours, respectively $\chi_{\rm FC}$ and $\chi_{\rm ZFC}$, are qualitatively reproduced by the RSB theory.
It holds $\chi_{\rm ZFC}=\beta(1- q_{\rm EA})$ and $\chi_{\rm FC}=\beta(1-\langle q \rangle)$, 
where $q_{\rm EA}$ is the Edwards-Anderson parameter, else called the self-overlap of a state with itself, 
and $\langle q \rangle$ denotes the average over the distribution $P(q)$. This early kind of experiments on spin-glasses provided modal and average values of the distribution.

 %--------------Figure ------------------------%
\begin{figure}[t!]
\centerline{\includegraphics[width=.7\textwidth]{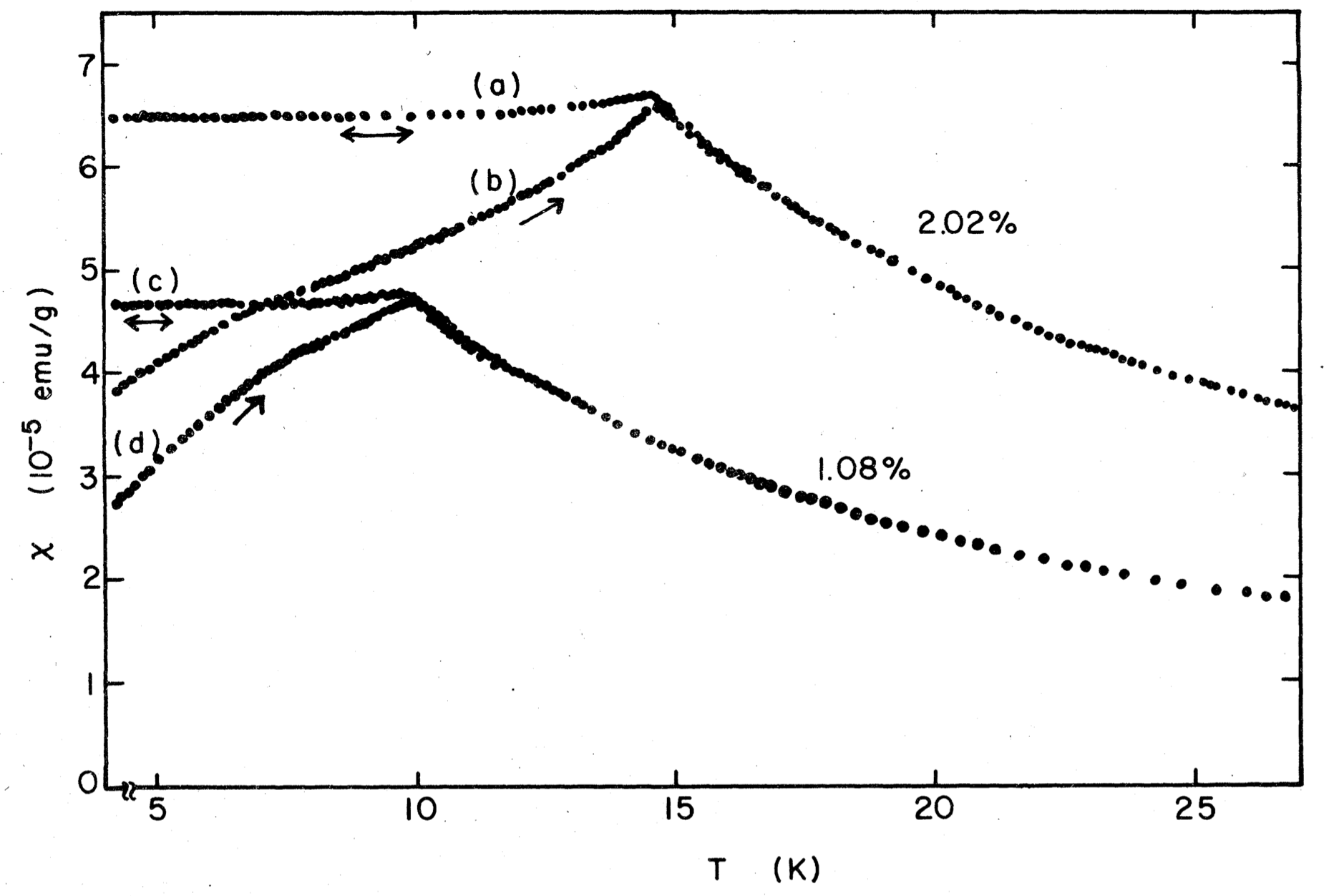}}
 \caption{Reproduced from Ref. \citeonline{Nagata79} with permission of the American Physical Society. Magnetic susceptibility of the spin-glass CuMn in the static limit. 
 Curves (b) and (d) are related to measurements taken after zero-field cooling and curves (a) and (c) to measurements after cooling in a field of $5.90G$.
 The two top curves (a) and (b) are the temperature dependece of the susceptibility of an alloy with $2.02\%$ of Manganese. The two lower curves (c) and (d)  are measured on a sample with $1.08\%$ of Mn. 
}
\label{fig-Nagata79}
\end{figure}
%----------------end figure -----------------%

How to measure the whole distribution?
In principle one needs to measure many configurations of spins $\{\sigma\}$ in time, in a well thermalized spin glass.
This task is very hard, because of the difficulty of measuring several atomic spins at once and because glassy systems hardly reach equilibrium.

Exploiting exactly the slow relaxation dynamics of glassy systems, an alternative procedure circumventing the direct measure of spins configurations was devised in 1998 by Franz, Mezard, Parisi and Peliti \cite{Franz98}. 
Under the assumption of stochastic stability they were able to prove that in aging out-of-equilibrium complex disordered systems, in the long time limit of both the waiting and the observation time the ratio between the response and the correlation comes out to be equivalent to the cumulative overlap distribution at equilibrium. 
In formulas, being the overlap $q$ the asymptotic  limit of the two-time self-correlation function, $$q=\lim_{t\to \infty}C(t,t'),$$ the long times limit of the fluctuation-dissipation ratio,
\begin{equation}
\tilde X(q) \equiv     \lim_{t,t'\to \infty} \frac{\chi(t,t')}{1-C(t,t')}  ,
\label{X_q}
\end{equation}
results to be equal to the cumulative (\ref{x_q}). Furthermore,  under those assumptions the integrated response function  depends on $t'$ and $t$ exclusively through the correlation function: $\chi(t,t')=\chi(C(t,t'))$.

\begin{figure}[t!]
\centerline{\includegraphics[width=1.03\textwidth]{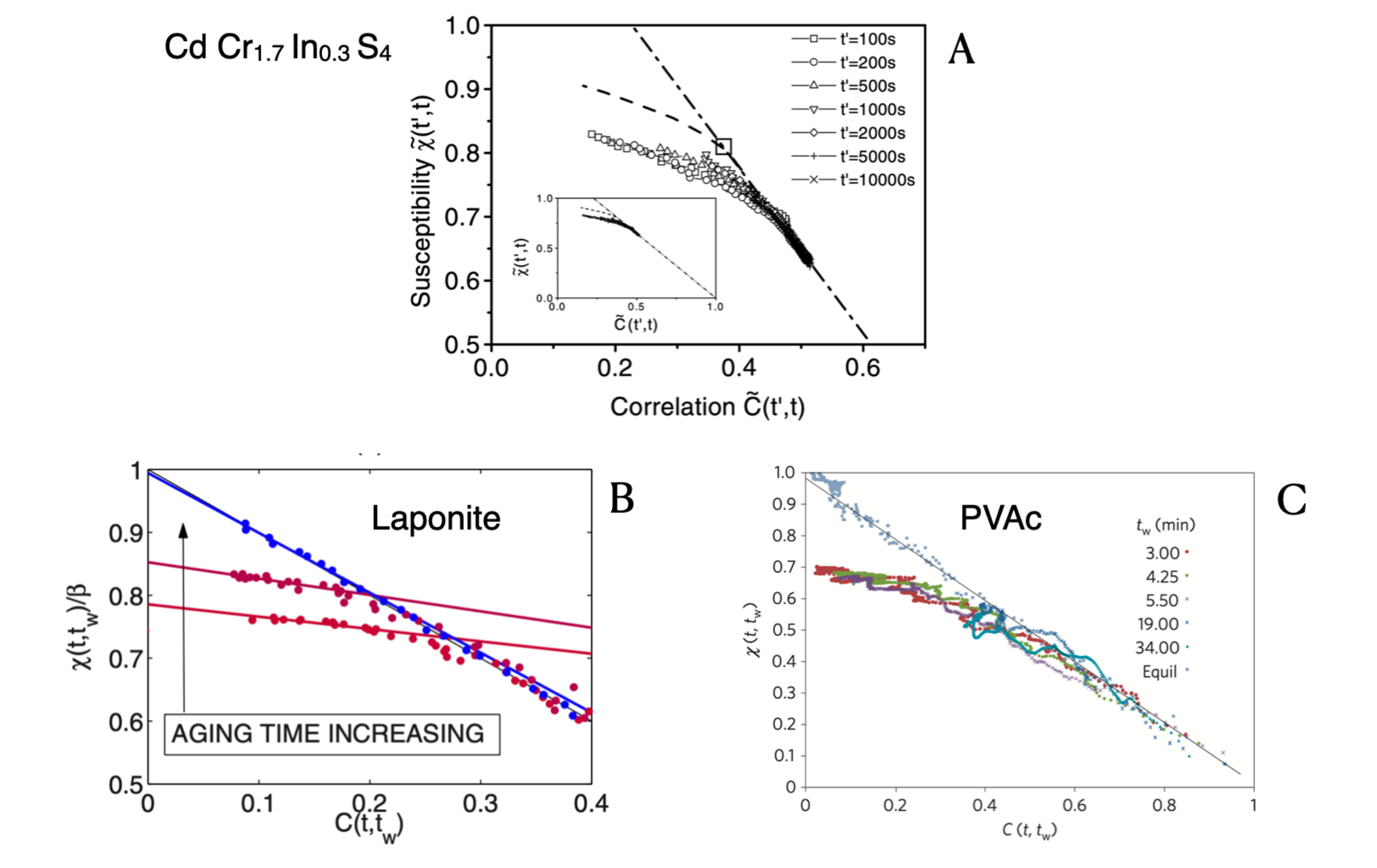}}
 \caption{Fluctuation-Dissipation ratio on ageing systems. (A) the ${\rm Cd Cr}_{1.7} {\rm In}_{0.3} \rm S_4$ spin-glass,  reproduced from Ref.~\citeonline{heri02} with permission from American Physical Society, (B) Laponite clay,  reproduced from Ref.~\citeonline{magg10} with permission from  American Physical Society, and (C) polivynil-acetate, reproduced from Ref.~\citeonline{oukr10} with permission from Nature Publishing Group.
}
\label{fig-FDRs}
\end{figure}

To measure such a fluctuation-dissipation ratio in real experiments, though, proved harder than expected.
Three impressive experiments were carried out in the last 20 years,\cite{heri02,heri04,magg10,oukr10} yielding the behaviours of response vs correlation behavior $\chi (C(t,t'))$  reproduced in figure \ref{fig-FDRs}. We briefly report and comment the outcomes.

 %--------------Figure ------------------------%
%----------------end figure -----------------%
The first experiment is on  samples of micrometric powder grains of an insulating  spin-glass  ${\rm Cd Cr}_{1.7} {\rm In}_{0.3} \rm S_4$\cite{heri02,heri04}. Very careful SQUID measurements of magnetic fluctuations allowed to provide the $ \chi(C)$ function of the correlation function $C(t,t')$ displayed in fig. \ref{fig-FDRs}-A.
The experiment is very accurate and deals with extremely weak thermodynamic fluctuations. Strict preacutions must be taken and kept for the whole duration of the experiment, but 
 very long measurements are required in order to acquire enough statistics. One day of measuremenys is required to provide data curves with long waiting times and, furthermore, the outcome does not  appear clean enough  to demostrate the equivalence of (\ref{x_q}) and (\ref{X_q}) in the infinite times limit.

In an experiment  on the  reorientational dynamics of Laponite disks\cite{magg10}, using the orientational correlation functions measured by depolarized dynamic light scattering and
 the corresponding response function via the electric field
induced birefringence, another kind of $ \chi (C)$ is obtained. In this case laponite is shown to undergo a glassy transition in density and not in temperature. 
At high packing density it provides the behavior of a (fragile) structural glass. Therefore, it is expected to be one step RSB-like in the asymptotic limit, as, actually shown in fig. \ref{fig-FDRs}-B, where  $ \chi(C)$  displays two slopes. In this experiment, though, as the waiting time $t'=t_w$ grows the FDR tends to the equilibrium ratio $1/T$ and the out-of-equilibrium counterpart of the cumulative overlap distribution appears to vanish.

Finally, the nanoscale polarization fluctuations and dielectric responses were measured in polyvinyl acetate (PVAc), \cite{oukr10} also known to be a fragile glass. PVAc displays aging at $T\simeq 0.98 \ T_g$ and at such temperature the system is expected to be one-step RSB, i. e., $ \chi(C)$ is expected to tend to a  function as the one in Fig. \ref{fig-FDRs}-C, whose derivative is a step function. Instead, according to the analysis with the available data a continuously bending $ \chi(C)$ is interpolated, hinting the occurrence of a continuous RSB, as in the proper spin-glass case, or hinting the occurrence of strong finite time pre-asymptotic effects.

To sum up, so far, through  stochastic stability in off-equilibrium dynamics  no $P(q)$ has  been experimentally demonstrated so far. 
Experimentally, indeed, things turn out to be different even from the most sophisticated numerical simulations, see [\citeonline{baityjesi17}] for a recent reference.

This is where photonics comes in: unlike single spins in amorphous magnets, or local degrees of freedom (e.g., density, orientational, polarization fluctuations) in structural glasses, single light mode intensities are degrees of freedom accessible to experimental measurements, at least partially. 
This {\em partiality} can be dealt with introducing a new overlap, the intensity fluctuation overlap, whose distribution is  equivalent to the Parisi overlap distribution (under given conditions that will be discussed in section \ref{SS-IFO}).

Before showing how a $P(q)$ can be acquired in experiments on random lasers and other photonic systems we very briefly recall how and to what extent a random laser is a complex disordered system possibly undergoing a transition to a RSB phase.

%%%%%%%%%%%%%%%%%%%%%%%%%%
\section{Random Lasers as Complex Disordered Systems} 

Random lasers are made of an optically active
medium and randomly placed scatterers\cite{Cao98,Cao99,Cao00,Anni04,Wiersma08,vanderMolen07,Tulek10,Andreasen11,Folli12,Viola18,Antenucci21,Gomes21,Elizer22} (sometimes both in one \cite{Cao99}). The first provides the gain, the latter provide the high refraction index  and the feedback mechanism needed
  to lead to amplification by stimulated emission.
As opposed to ordered standard multimode lasers, random lasers do not require complicated construction and rigid optical alignment, 
 have omni-directional emission and
high operational flexibility. They give rise to a number of promising applications
 in the field of speckle-free imaging \cite{Redding12,Barredo-Zuriarrain17}, granular matter \cite{Folli12,Folli13}, remote sensing \cite{Ignesti16,Xu17,Gomes21}, medical diagnostics and biomedical imaging \cite{Polson04,Song10,Lahoz15,Wang17,Gomes21}, optical amplification  and optoelectronic devices \cite{Lin12,Liao16,Gomes21}. 

Random lasers emission spectra above a pump threshold may show multiple sub-nanometer spectral peaks, 
 as well as 
single narrow curves with  5-10 nm width. Depending on the material, and its optical and scattering properties, random spectral fluctuations between different pumping shots (i. e., different realizations of the same random laser) may or may not vary significantly.
A wide variety of  spectral features is reported \cite{Turitsyn10,Leonetti11,Folli13,Baudouin13,Antenucci21}, depending on material compounds and experimental setups. Random lasers can be built in very different ways, can be both solid or liquid, can be 2D or 3D, the optically active material can be confined or spread all over the volume. 
Moreover, random lasers are, usually, open systems where light can propagate in any direction rather than being confined between well specific boundaries (mirrors) as in standard lasers and the emission acquisition can  be on the whole solid angle.

\subsection{The leading model Hamiltonian}

\begin{figure}[t!]
\centerline{\includegraphics[width=.8\textwidth]{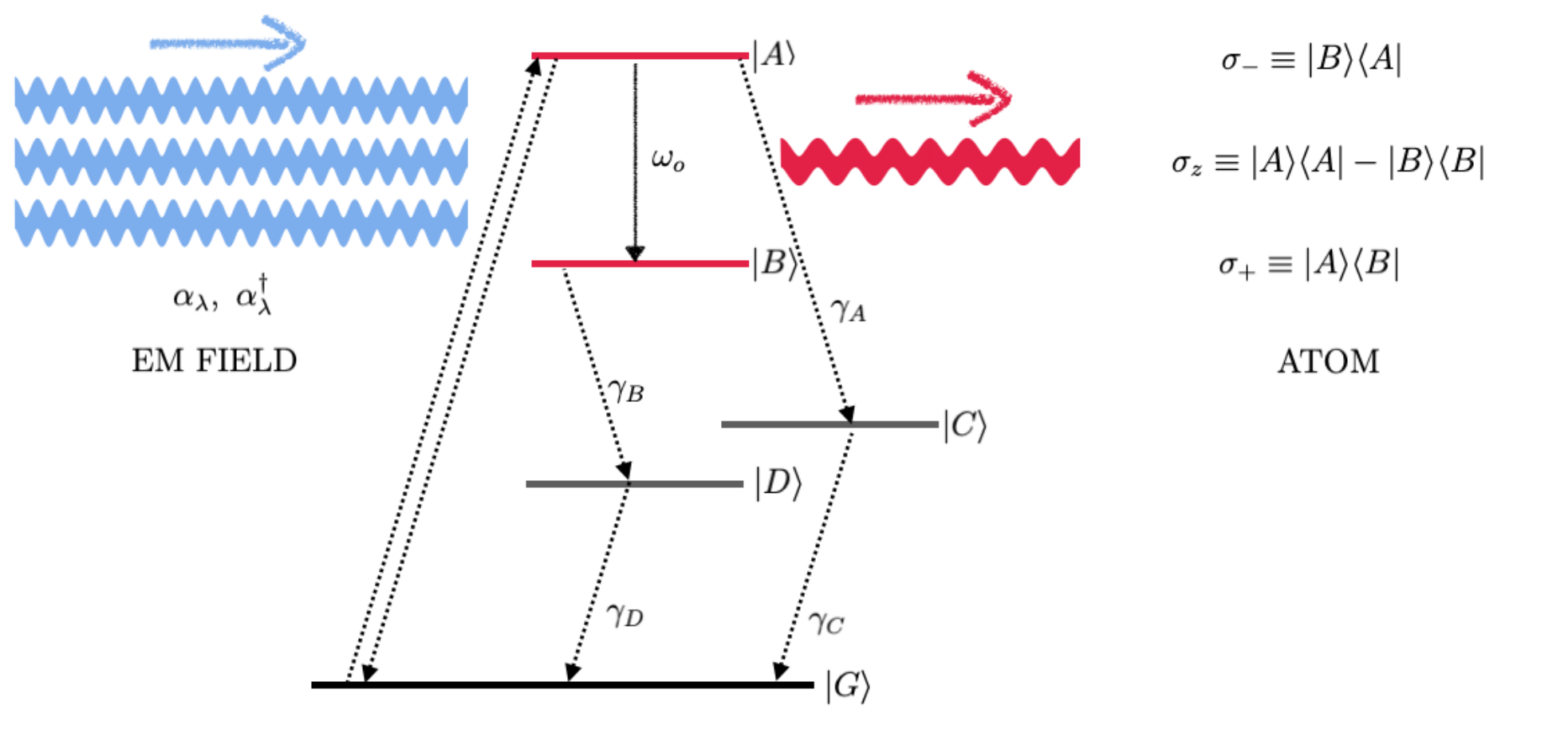}}
 \caption{Pictorial representation of the light-matter interaction. The atom here illustrated  has five states denoted by letters  $|A\rangle$, $|B\rangle$, $|C\rangle$, $|D\rangle$ and $|G\rangle$, the ground state. $\gamma_x$, with $x=A,B,C,D$ are the decay rate of the excited states to a state of lower energy.
 The energy gap between $|A\rangle$ and $|B\rangle$ is $\hbar \omega_o$ where $\omega_o$ lies in the optical frequency spectrum and contributes to the laser in the regime of high external pumping, when many atoms coherently emit  at once. 
}
\label{fig-JC}
\end{figure}

Since the original proposal that a random laser might be described by means of a spin-glass-like Hamiltonian,\cite{Angelani06a,Angelani06b} various derivations  have been put forward \cite{Con11,Ant15c,Ant16,Ant16b} of the leading model 
\begin{equation}
    \mathcal H[\bm a] = -\sum_{k_1,k_2}  J^{(2)}_{k_1k_2}a^*_{k_1}\ a_{k_2} - \sum_{k_1,k_2,k_3k_4} J^{(4)}_{k_1k_2k_3k_4} a_{k_1}a^*_{k_2}a_{k_3}a^*_{k_4}  + \mbox{c.c.},
\label{d-leading-hamiltonian}
\end{equation}
where $a$'s are complex numbers denoting the complex amplitudes of the light modes and the $J$'s denote quenched random mode-couplings.
The most fundamental derivation is the construction starting from the light-matter interaction between the atoms or molecules of the gain medium, thus displaying an optical gap, and the  electromagnetic (e.m.) field of the light.\cite{Ant15c} We briefly skecth the basic steps moving from a quantum description of the operator dynamics to a classical Hamiltonian theory for the 
stationary regime of light amplification by stimulated emission in a disordered medium.
The quantum stochastic differential equations describing the interaction of an atom with an optical gap $\omega_o$ between two levels $|A\rangle$ and $|B\rangle$, see figure \ref{fig-JC}, with the electromagnetic field of frequency $\omega_\lambda$, represented by the creation and annihilation operators $\alpha$ and $\alpha^\dagger$, are the Jaynes-Cummings equations\cite{jay63}
       \begin{eqnarray}
         \dot{\alpha}_{\lambda} & =&  - \imath \omega_{\lambda} a_{\lambda} - \sum_{\mu} \gamma_{\lambda \mu} \alpha_{\mu} + %
         \int d \bm{r} g^{\dagger}_{\lambda}(\bm{r}) \  \sigma_{-}(\bm{r}) + F_{\lambda} \label{eq:dyn_mode} \\
         \dot{\sigma}_{-}(\bm{r}) & = & -(\gamma_{\bot} + \imath \omega_o) \sigma_{-}(\bm{r}) + %
         2 \sum_{\lambda} g_{\lambda}(\bm{r})\ \sigma_z(\bm{r})\  \alpha_{\lambda} + F_{-}(\bm{r}) \label{eq:sigma_} \\
         \dot{\sigma}_{z}(\bm{r}) & = & \gamma_{\parallel} \left(S \rho(\bm{r}) - \sigma_z(\bm{r})\right) - %
         \sum_{\lambda} \left(g^{\dagger}_{\lambda}(\bm{r})\ \alpha^{\dagger}_\lambda\sigma_{-}(\bm{r}) + \text{h.c.} \right) + %
         F_{z}(\bm{r}) , \label{eq:sigma_z} 
         \end{eqnarray}
where $\sigma_z(\bm r) \equiv |A\rangle \langle A| - |B\rangle \langle B|$ is the population inversion operator, $\sigma_-(\bm  r)\equiv   |B\rangle \langle A|$ is the lowering operator and $\sigma_+(\bm  r)\equiv   |A\rangle \langle B|$ the raising operator.
The coefficients $g_\lambda(\bm r)$ are the atom-field coupling constants, $\gamma_{\mu\nu}$ is the damping matrix and it is associated to the fact that the cavity is open in random lasers and radiative modes are there, as well. The coefficients $\gamma_{\bot}$,  $\gamma_{\parallel}$, are, respectively, the polarization decay rate and the population inversion rate ($\gamma_\parallel \equiv (\gamma_A+\gamma_B)/2$, see figure \ref{fig-JC}).
The atomic density is $\rho(\bm r)$ and $S$
represents   the  intensity of the external optical pumping.

 %--------------Figure ------------------------%
%----------------end figure -----------------%

Noise terms $F_\lambda$, $F_-(\bm r)$ and $F_z(\bm r)$ are there, as well, and they pertain to  different stochastic  phenomena. The term $F_\lambda$ is due to the presence of radiative modes in open cavities and the effetive interation between those and the modes inside the cavity.
The atomic noise terms $\mathcal F_-$ and $\mathcal F_z$
 arise because of the intercation between the e.m. field and the optically active  medium.

Applying perturbation theory the dependence of the atomic operators $\sigma_-$ and $\sigma_z$ on the field operator $\alpha_\lambda$ can be worked out, leading to a single equation for  coupled $\alpha$'s.
Degrading from quantum operators to complex numbers we are, eventually, left with a single stochastic differential equations for the electromagnetic field modes $\alpha$ in the cavity. 
These modes, at frequencies $\omega_\lambda$, are called ``cold cavity'' or ``passive'' modes because they are 
intrinsic of the system and not activated by any 
 external pumping. 
 In presence of external excitation and above the lasing threshold 
the evolution in the lasing regime is better expressed in the basis $\{E_k(\bm r)\}$ of the so-called slow amplitude modes. A slow amplitude mode $a_k$ is a mode that, for  long enough time, displays a harmonic oscilating behavior at some given angular frequency $\omega_k$, so that the overall electromagnetic field can be written as
\begin{equation}
    \label{E_r_t}
    E(\bm r, t) = \sum_{k=1}^N a_k(t)\ 
    E_k(\bm r) \ e^{\imath \omega_k t} \ \ \ + \ \ \mbox{c.c.}
\end{equation}
for a system with $N$ modes.
The relationship between passive and slow amplitude modes is not unique and can always be expressed in the form
\begin{equation}
\alpha_\lambda(t) = \sum_{k} M_{\lambda k} \ a_k(t) \ e^{\imath \omega_k t}.
    \label{eq-cold-slow}
\end{equation}
For a simple case where the relationship between passive and lasing modes can be worked out exactly at the quantum level one can see, e.g., Refs. [\citeonline{Eremeev11,Marruzzo15th}].
It is called slow amplitude approximation because the dynamics of $a_k(t)$  is much slower than the one of the  mode oscillation $e^{\imath \omega_k t}$, so that the Fourier transform of $a_k(t) \ e^{\imath \omega_k t}$ tends to a Dirac delta $\delta(\omega-\omega_k)$.

Carrying out this transformation eventually leads to a stochastic differential equation of the potential kind for the slow amplitudes:
\begin{eqnarray}
\dot a_k(t) = -\frac{\partial \mathcal H[\bm a]}{\partial a^*_k} + F_k(t)
\label{eq-eds-amplitudes}
\end{eqnarray}
where the noise $F_k(t)$ can be taken as white noise and the Hamiltonian $\mathcal H[\bm a]$ turns out to be 
\begin{equation}
    \mathcal H[\bm a] = -\sum_{\bm k \ | \ {\rm FMC(\bm k)}} J^{(2)}_{k_1k_2}a^*_{k_1}\ a_{k_2} - \sum_{\bm k \ | \ {\rm FMC(\bm k)}} J^{(4)}_{k_1k_2k_3k_4} a_{k_1}a^*_{k_2}a_{k_3}a^*_{k_4}  + \mbox{c.c.}. 
    \label{eq-Ham-leading}
\end{equation}
In the above Hamiltonian definition a constraint is imposed on the frequencies induced by the slow amplitude condition, that we call the Frequency Matching Condition (FMC). 
For two and four modes it reads, respectively,
\begin{eqnarray}
   && \left|\omega_{k_1}-\omega_{k_2}\right| \lesssim \gamma 
    \label{eq-fmc2}
    \\
    &&  \left|\omega_{k_1}-\omega_{k_2}+\omega_{k_3}-\omega_{k_4}\right| \lesssim \gamma  , 
    \label{eq-fmc4}
\end{eqnarray}
where $\gamma$ is the finite linewidth of the modes. This can only be derived  in a quantum theoretical approach,\cite{Eremeev11} whereas, for what concerns the classical approach, we can include $\gamma$ as a parameter coherent with experimental observations.
In equation (\ref{eq-Ham-leading}) two effective coupling terms appear: a two- and a four-mode coupling, whose expressions are derived as the Hamiltonian is built. In the slow amplitude mode basis they depend on the spatial intersection of the eigenfunctions $E_k(\bm r)$ of the modes modulated by the spatial profile of, respectively, the linear $\chi^{(1)}$ and  the nonlinear $\chi^{(3)}$ susceptibility of the random medium:
\begin{eqnarray}
J_{k_1k_2}&\propto&  \int d \bm r \ E_{k_1}(\bm r) \ E_{k_2}(\bm r) \  \chi^{(1)}(\bm r | \omega_{k_1}, \omega_{k_2})
\label{def-J2}
\\
J_{k_1k_2k_3k_4}&\propto&
\int d \bm r \ E_{k_1}(\bm r) \ E_{k_2}(\bm r) \ E_{k_3}(\bm r) \ E_{k_4}(\bm r) \ \chi^{(3)}(\bm r | \omega_{k_1}, \omega_{k_2}, \omega_{k_3}, \omega_{k_4})
\label{def-J4}
\end{eqnarray}

One last fundamental ingredient of modes dynamics in a lasing material to be combined   in the formulation of the Hamiltonian dynamics is gain saturation. 

Gain saturation is fundamental to  have a stationary solution at all in the systems under study, that might  be represented as a potential, equilibrium-like, solution to the stochastic equations. 
The most ``energetic'' phenomenon that can occur in a system of atoms in a laser is when  they all are in their excited optical level and they all emit a photon at once.
As such an event were to  occur, the atoms emission would be  soon afterwards depleted because they have to invert their population and that takes time. Therefore, the gain - that is the capability of the material of amplifying light -  saturates at a certain level $E_{\rm sat}$. Its behavior as a function of the total energy $\mathcal E$ pumped into the system, is usually modeled as \cite{Gordon02,Ant16} 

$$g(\mathcal E) = \frac{g_0}{1+\frac{\mathcal E}{E_{\rm sat}}}.$$ 

\begin{figure}[t!]
\centerline{\includegraphics[width=.9\textwidth]{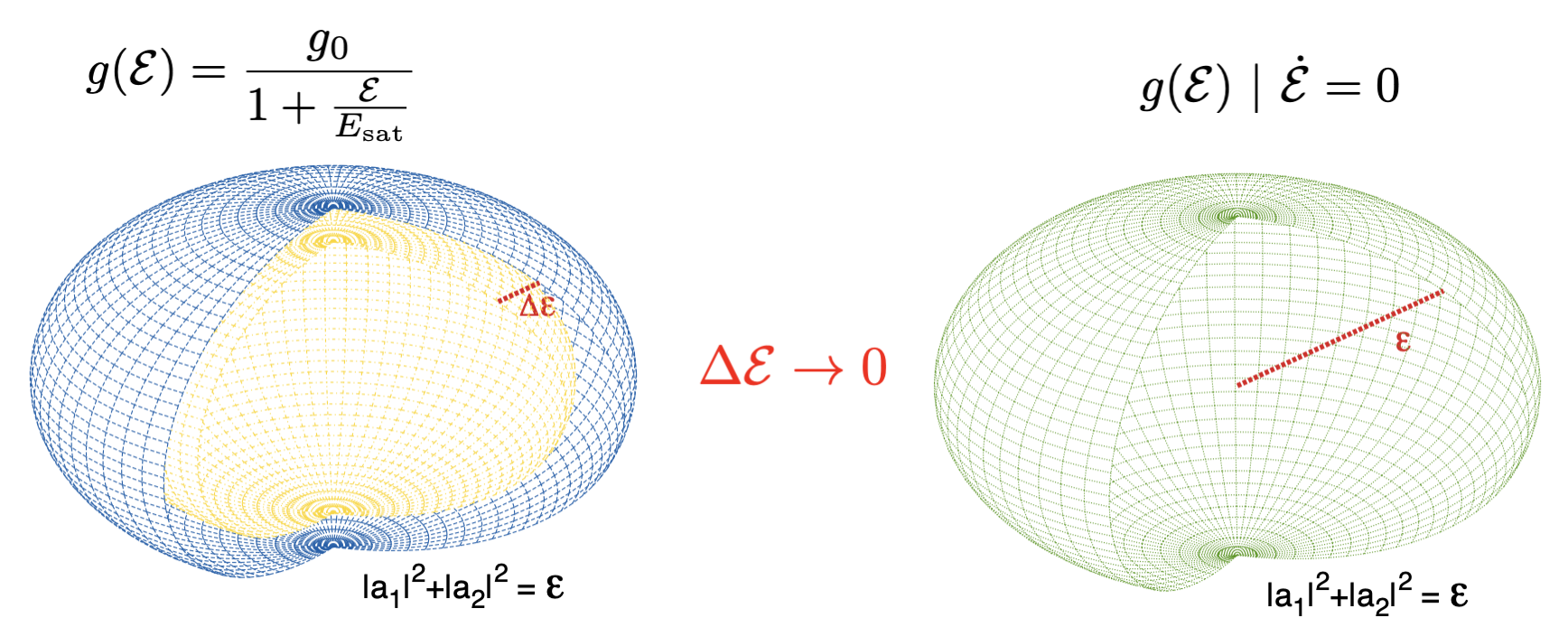}}
 \caption{A sketch of the spherical approximation for the gain saturation.
}
\label{fig-gs}
\end{figure}

Even though the external pumping continues  the amount of lasing will not increase indefinitely, no matter how strong  the pumping is , but it will reach a stationary regime. 
The overall energy $\mathcal E$ shared by the modes in the cavity slowly fluctuates between an upper and a lower bound, but its change is much slower than the one of the single mode intensities.
As a further approximation, then, in the spin-glass theory for random lasers we assume that the dynamics of the total energy of $N$ modes is so slow to be considered as a constant with respect to the dynamics of the single photon emissions at any time $t$, i.e.,
\begin{equation}
    \mathcal E = \sum_k |a_k(t)|^2 = \epsilon \ N = \mbox{const}.
    \label{eq-spherical}
\end{equation} 
In figure \ref{fig-gs} we pictorially sketch this approximation in a two-mode case. 
From the point of view of the phasors $a_k(t)$, whose dynamics is governed by (\ref{eq-eds-amplitudes}),  Eq. (\ref{eq-spherical}) is a global contraint for the dynamics of a configuration to be on the $2N$ hyper-sphere of radius $\sqrt{\epsilon N}$.
This is the reason why this kind of spins are refereed to as ``spherical'' in statistical mechanics literature \cite{Crisanti92}. We will, therefore, call the leading model (\ref{eq-Ham-leading}) a complex spherical $2$+$4$-spin spin-glass model. 

With such a global constraint
 the stationary regime can be described as if the system were at equilibrium at an effective  temperature (a ``photonic'' temperature $T_{\rm ph}$) related to the ratio $\mathcal P$ between the external pumping rate and the spontaneous emission rate. The latter is proportional to the real heat bath temperature $T$, whereas the first one is proportional to the energy (\ref{eq-spherical}) stored into the photonic system. 
In a formula,  the role of the temperature driving the lasing transition is played by the rescaled temperature
\begin{equation}
T_{\rm ph} = \frac{T}{\epsilon^2} = \frac{1}{\mathcal P^2} = \frac{1}{\beta_{\rm ph}}.
\label{eq-temp}
\end{equation}

 %--------------Figure ------------------------%
%----------------end figure -----------------%

\subsection{An analytic solution in the narrowband approximation}
\label{SS-NB}

The spin-glass model (\ref{eq-Ham-leading}) in the limit of narrow bandwidth of the random laser spectrum tends to the generalization of the fully connected spherical  $2$+$4$-spin spin-glass model\cite{Crisanti04b,Crisanti06,Crisanti07,Crisanti13} to complex spins.
In the narrowband approximation\cite{Chen94,Gordon02} all resonances have frequencies  so close to each other that their difference is of the order of the linewidth $\gamma$, so that the conditions (\ref{eq-fmc2}) and (\ref{eq-fmc4}) are always satisfied. 
Under the further assumption that all modes have a  spatial extension of the order of the whole volume of the  lasing material, Eqs. (\ref{def-J2}) and (\ref{def-J4}) imply that the interaction network can be taken as fully connected.

The fully connected limit has been extensively studied in the thermodynamic limit in Refs. [\citeonline{Ant15b,Ant15c,Ant16,Ant16b}] where the couplings (\ref{def-J2}) and (\ref{def-J4})  are chosen to be independent identically distributed gaussian variables of given mean $\bar J/N^\delta$ and variance $\sigma_{J}^2/N^\delta$. 
The exponent is $\delta=1$ in the two-body term and $\delta=3$ in the four body term. In this way the magnitude of each coupling decreseas as the number of coupling of a single mode and the Hamiltonian (\ref{d-leading-hamiltonian}) on the fully connected graph of interaction if always extensive.
%, $\mathcal O(N)$.

\begin{figure}[t!]
\centerline{\includegraphics[width=.9\textwidth]{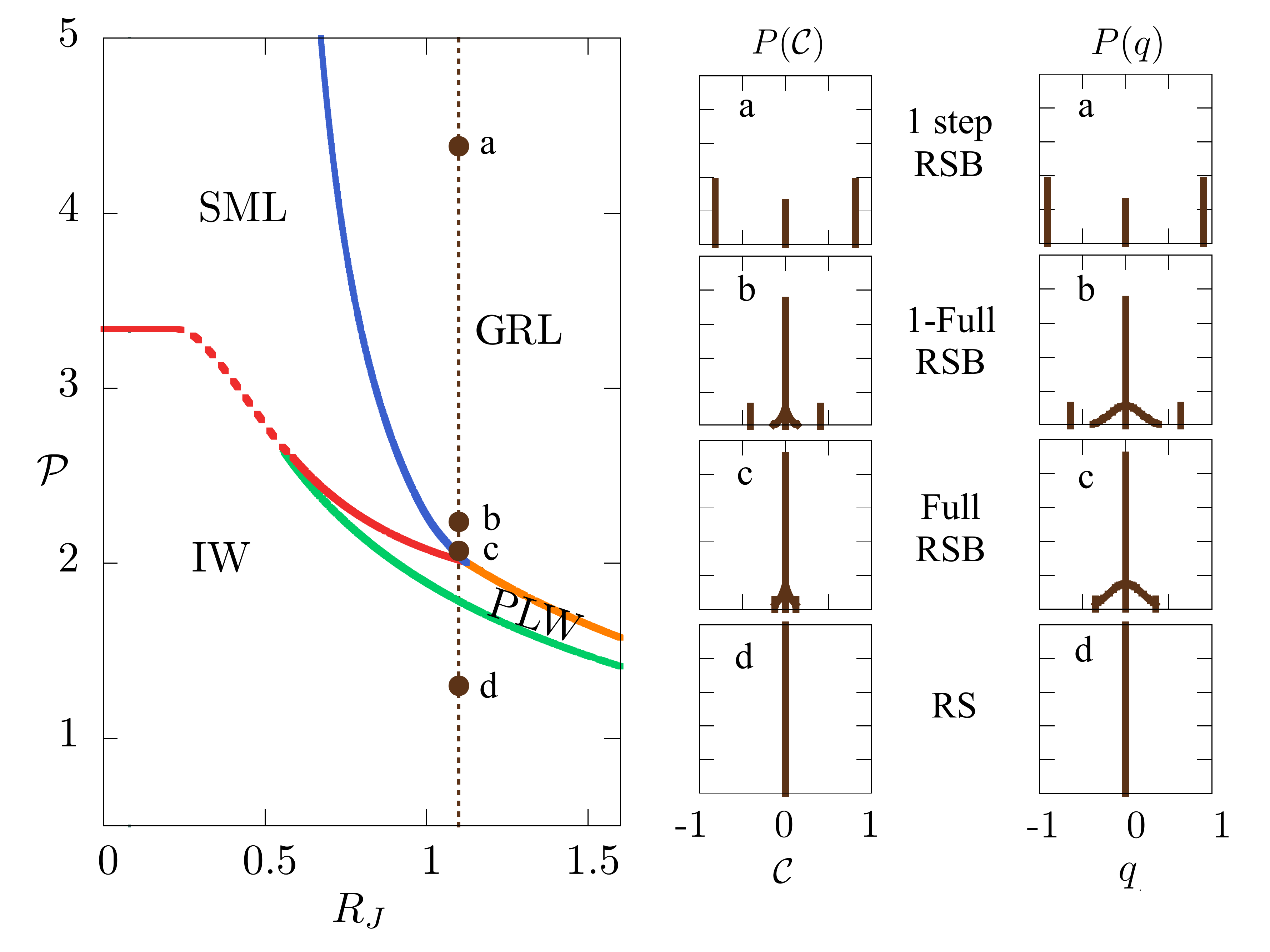}}
 \caption{An instance of a phase diagram of a random laser model (\ref{d-leading-hamiltonian}) in the narrowband approximation (left). Four thermodynamic phases can occur depending on the degree of randomness $R_J$: Incoherent Wave (IW) at low $\mathcal P$ and low disorder $R_J$, Standard Mode locking (SLM) at low $R_J$ and high pumping $\mathcal P$,  Phase-Locking Wave (PLW) in a range of disorder $R_J$  large but not extremely large, Glassy Random Laser (GRL)  above a certain pumping threshold for large $R_J$. The latter is the regime of broken replica symmetry and needs to be analyzed by means of Parisi theory for complex disordered systems.  
 The dotted lines are first order transition lines, that is, above and below those dotted lines a region of phase separation is expected (spinodal lines, and how to computed them, are reported in Ref. \citeonline{Ant15c}).
 On the right the IFO and Parisi overlap distributions $P(\mathcal C)$ and $P(q)$
 are displayed in four points of the phase diagram along a line of increasing pumping for a given degree of disorder. For low pumping the IW regime is described by a replica symmetric solution. Increasing $\mathcal P$ the system becomes a GRL with a Full RSB. Increasing further $\mathcal P$ the laser moves to a 1-Full RSB phase and, eventually, for very large pumping, the  behavior of the GRL is described by  a one step RSB phase. 
 Repoduced from Ref.~\citeonline{Ant15a} with permission from Nature Publishing Group.}
\label{fig-phdi}
\end{figure}

The ratio $R_J\equiv \sigma_J/\bar J$ between mean square displacement  and mean value represents the degree of disorder of the system.\footnote{The model can be implemented with two different degrees of randomness for the $2$- and the $4$-body interactions \cite{Ant15c} but for the sake of the presentation we consider them equal to each other.} {\em Glassy} random lasers, i.e., random lasers displaying anomalous shot-to-shot fluctuations of the emission spectra, will generally have a large $R_J$ (even infinite if we take $\bar J=0$). In Figure \ref{fig-phdi} we reproduce  a typical phase diagram in the pumping rate and degree of disorder.
Four thermodynamic phases can occur depending on the degree of randomness $R_J$. As the disorder is small (or none) the model (\ref{d-leading-hamiltonian}) displays   an incoherent wave regime at low pumping and a Standard Mode-Locking laser regime at high pumping $\mathcal P$. This regime also represents Random Lasers with no glassy features, that is,  lasers with random resonances in the spectrum but no anomalous flutucations from shot to shot (the resonances are random but they are always the same). As randomness in the mode-coupling increases a Phase-Locking Wave regime is predicted to occur, that is a regime where the phases of the modes are locked while no resonance occurs in the mode intensities. This phase vanishes as $R_J\to\infty$. Eventually, for large enough randomness a Glassy Random Laser occurs above a certain pumping threshold.   This is the regime where the spin-glass theory of multistate systems  is necessary in order to deal with the complexity of the spectral behavior. 

In the highly (quenched) disordered region 
the order parameter  identifying the random lasing threshold is the probability distribution of the overlap
\begin{equation}
    \label{d-q-Parisi}
q_{\rm{ab}}= \Re \left[\frac{1}{N}\sum_{k=1}^N \bar a^{(\rm a)}_k a^{(\rm b)}_k\right].
\end{equation}
between any two replicas a and b.

 Some instances of RSB solution in different points of the phase diagram, and the relative $P(q)$'s are reported in Figure \ref{fig-phdi} (the Parisi distributions are displayed in the rightmost column). 
 For low pumping the incoherent wave regime is described by a replica symmetric solution. 
 Increasing the pumping the system undergoes a transition to a Glassy Random Laser with  infinite breakings of the replica symmetry (Full RSB). 
 Increasing further $\mathcal P$ the laser moves to a 1-Full RSB phase and, eventually, for very large pumping, where the pairwise (i.e., linear) contribution to (\ref{d-leading-hamiltonian}) plays no role anymore, a one step RSB phase correctly describes the thermodynamic behavior fo the Glassy Random Laser.

 %--------------Figure ------------------------%
%----------------end figure -----------------%

To measure such an overlap one has to access real and imaginary parts of the complex amplitude of each light mode $a_k$ in each replica. Else said the intensity and the phase:
\begin{eqnarray}
I_k^{(\rm a)}&=& \left|a_k^{(\rm a) }\right|^2
\label{d-intensity}
\\
\phi_k^{(\rm a)}&=&\mbox{arg}\left(a_k^{(\rm a) }\right).
\label{d-phase}
\end{eqnarray}
As will be discussed in a while, even though the intensities can easily be measured  from the emission spectra, the phases in random lasers are  hardly accessible experimentally.
To overcome such a hindrance it is possible to define an overlap between replica involving intensities alone. Or, better, involving their fluctuations.

\subsection{The intensity fluctuation overlap (IFO)}
\label{SS-IFO}
Let us define the fluctuation between the intensity  $I_k^{(\rm a)}$ of a single resonance (a single mode $k$) in the spectrum of a single replica $($a$)$ and its average at equilibrium $\langle I_k^{(\rm a)} \rangle$:
\begin{equation}
    \Delta_k^{(\rm a)} =\frac{I_k^{(\rm a)}-\langle I_k^{(\rm a)} \rangle }{2\sqrt 2\ \epsilon}=\frac{I_k^{(\rm a)}-\langle I_k^{(\rm a)}\rangle }{2\sqrt{ 2  T}},
\end{equation}
where the normalization factor $\sqrt{8}$ is a pedantry and simply depends on the coefficients in the definition of 
(\ref{d-leading-hamiltonian}).

We can, then, introduce the overlap between the intensity fluctuations of two replicas $\rm a$ and $\rm b$:
\begin{equation}
\mathcal C_{\rm ab}\equiv \frac{1}{N} \sum_{k=1}^N \Delta_k^{(\rm a)}\Delta_k^{(\rm b)} .
    \label{d-IFO}
\end{equation}

In the fully connected model with large disorder the distribution of the Intensity Fluctuation Overlap (IFO) is proved to be equivalent to the one of the Parisi overlap (\ref{d-q-Parisi}) squared. Indeed, for each couple of replicas the identity holds:
\begin{equation}
    \label{equivalenza}
    \mathcal C_{\rm a b}= q_{\rm ab}^2 \qquad , \qquad  \mbox{a}\neq \mbox{b}
\end{equation}

In Figure \ref{fig-phdi} we show some examples of both distributions as the system undergoes a phase transition from an incoherent wave regime to a lasing regime that is random {\em and} glassy.

\subsection{Mode-locked random laser theory and numerical simulations}
Moving to more realistic random lasers one should relax the narrow-band assumption. Indeed, the optically active materials composing random lasers have, usually, a wide emission spectrum in which many resonances take place above the lasing threshold. 
That is, a more realistic system is modeled by equation (\ref{eq-Ham-leading}), implementing the conditions (\ref{eq-fmc2}), (\ref{eq-fmc4}). 
Such conditions imply a dilution of the interaction network of $\mathcal O(N)$ \cite{Marruzzo15th,Marruzzo18} and induce some kind of metrics in the space of the frequencies. \footnote{
 A model with $4$-body interaction, that represents the fundamental non-linear ingredient to yield lasing at high pumping (low temperature) will, therefore, display $\mathcal O(N^3)$ mode-coupling terms in the Hamiltonian (if  the modes are spatially extended to a finite fraction of the volume of the lasing material).
This also implies that the $\delta$ exponent introduced in section \ref{SS-NB} in the average and variance of the random coupling distribution will be descreased by one. 
The $2$-body term will have a sparse structure ($\delta=0$: each mode has a finite number of connection, whatever $N$), whereas it will be $\delta=2$ for the $4$-mode couplings.}

Even though the $2$-mode terms are important for the occurrence of a RSB solution of the full kind, in this section we will focus on the $4$-mode contribution alone, that is responsible for the onset of a glassy lasing regime, though as a one step RSB solution.
This means that in the lasing regime one should expect  a $P(q)$, or $P(C)$, like in figure \ref{fig-phdi}-d and \ref{fig-phdi}-a: a single central peak at high temperature that, as the lasing threshold is overcome, also displays two more side peaks at low $T$.

To analytically solve a diluted (that is not sparse, nor fully connected) system  with a deterministic prescription  is still an unsolved problem in replica theory but one can resort to Monte Carlo numerical simulations. \footnote{We also stress that to use a sparse approximation in this case would not allow to use the cavity method because the variables are not locally confined, the ``spins'' are not $|s|^2=1$ as the Ising, the XY \cite{Marruzzo15,Marruzzo16} or the Heisenberg spins. 
Instead light mode amplitudes only have an overall constraint (\ref{eq-spherical}) and locally one or a few of them might condensate the whole power at disposal, $|a|^2= \mathcal O(N)$, leaving nothing for all the others. This is a case where the equivalence of the canonical and the microcanonical ensembles break down and cannot be studied in the present form.}

\begin{figure}[t!]
\centerline{\includegraphics[width=.49\textwidth]{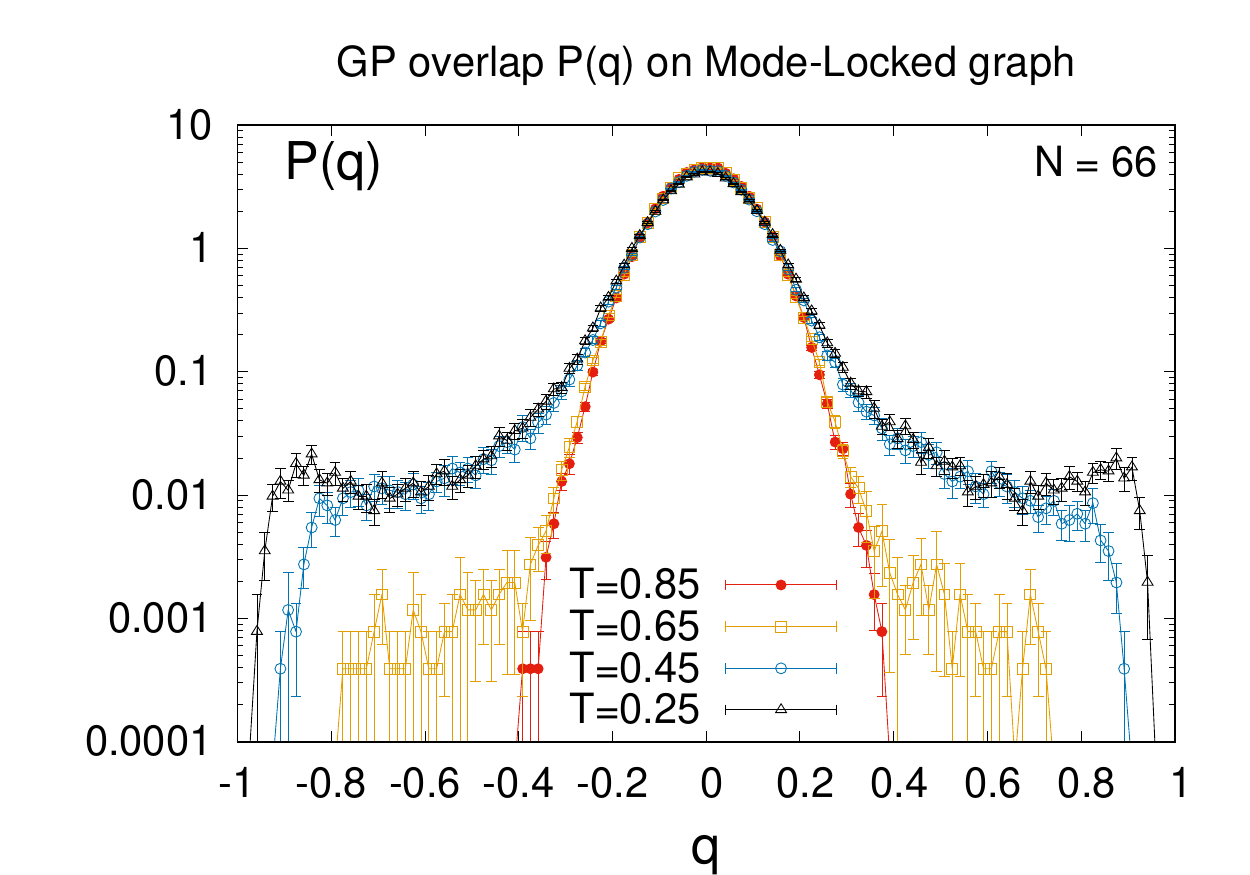}
\includegraphics[width=.49\textwidth]{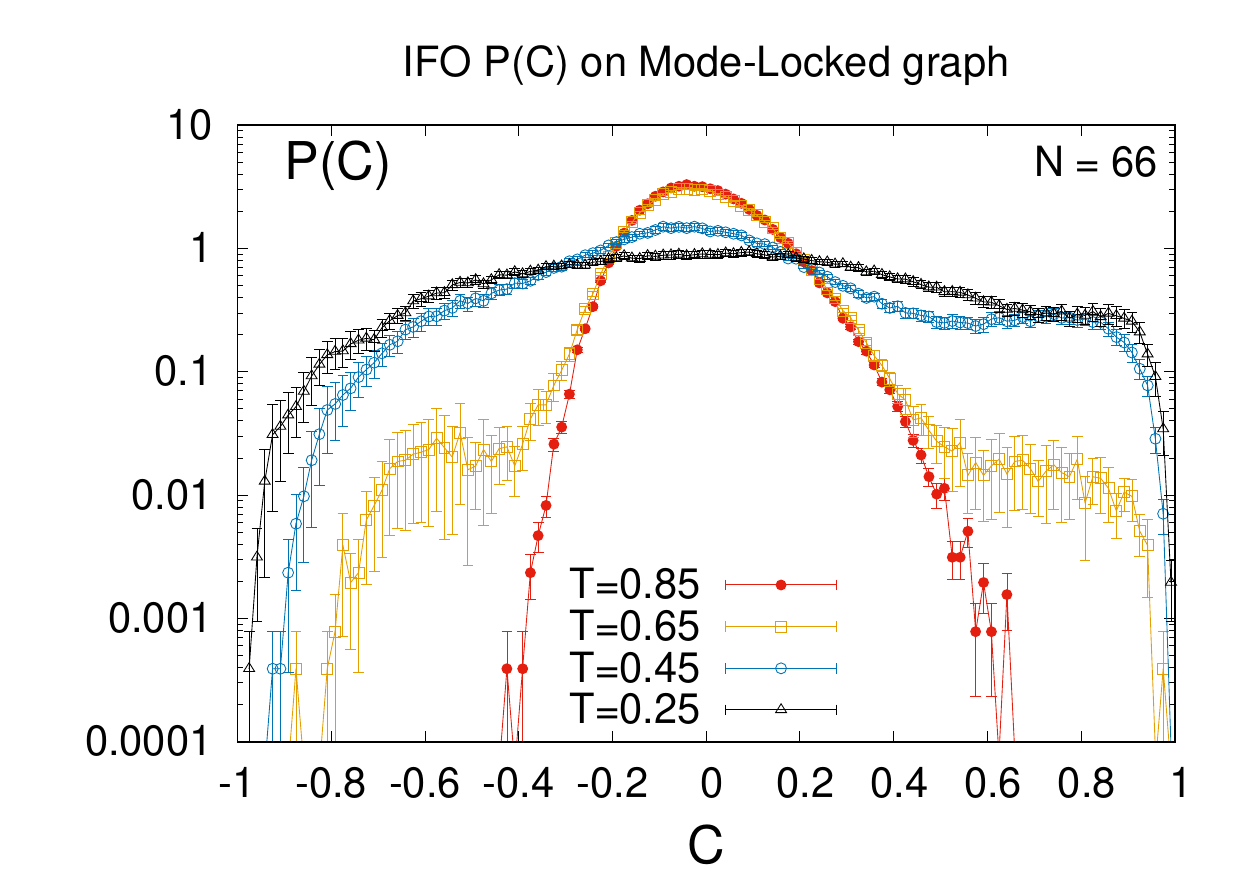}}
 \caption{Left: average Parisi overlap distribution $P(q)$ from numerical simulation of the dynamics of  the $4$-spin complex spherical spin model with $\bar J=0$ (``infinite'' degree of randomness $R_J$), $N=66$ modes at four different temperatures across the critical point, $T_c=0.61(3)$\cite{Niedda22}.  Right: average distribution $P(\mathcal C)$ of the Intensity fluctuation overlap, Eq. (\ref{d-IFO}), for the same system. In both plots, at each temperature, distributions are averaged over $100$ disordered samples. All samples are thermalized, that is the configurations are all at equilibrium and the time average coincides with the ensemble average.  
}
\label{fig-Pq-PC-num}
\end{figure}

In figure \ref{fig-Pq-PC-num} we report instances of average overlap distributions at various temperatures $T$ for a random laser model of $N=66$ on a mode-locked graph obtained simulating different replicas with the Exchange Monte Carlo, or ``Parallel Tempering'' algorithm. This exploits the properties of detailed balance in a  Markov chain dynamics of the system of spins through different parallel thermal baths in order to drastically reduce the thermalization time in the spin-glass phase. 
The implementation of the dynamics, moreover, is carried out on GPU's so that both the just mentioned parallel dynamics and the computation of every single energy update can be spread simultaneously on several threads.
This is a technical but crucial point in order to have reasonable simulation times in a system that is Non-Deterministic Polynomial Complete, has continuous variables and a number of interactions growing like $N^3$. We notice that the IFO distributions, that in the fully connected case are equivalent to the Parisi distributions, on the mode-locked graph appear to have much stronger finite size effects, as analyzed in Ref. \citeonline{Niedda22}. On the other hand their signal of a transition to a multi-state phase of nontrivial correlations is much more evident than with the $P(q)$, that one has to look in the log scale to appreciate the occurrence of side peaks.

We stress that, even if diluted with the FMC, Eq. (\ref{eq-fmc4}), the thermodynamic solution is expected to be a mean-field one, as in the case of the fully connected model, as confirmed by recent accurate numerical results \cite{Niedda22}.

 %--------------Figure ------------------------%
%----------------end figure -----------------%

When moving from analytical and numerical results to real experiments 
 different aspects  have to be taken into account. 
 As in the numerical simulations, also in  experiments the number of modes is finite. However, in experiments on glassy random lasers it is rather difficult to scale the system studying a controlled trend in $N$ and perform finite size scaling as in numerical simulations \cite{Gradenigo20}. 
Moreover,  also the mode resolution is finite and 
we do not know much about the spatial extension of the modes and, therefore, about the interaction graph, nor about  the magnitude and sign of the couplings. 
As anticipated, when the IFO are introduced in experiments only the mode intensity is acquired and not the mode phase.
Moreover,
  one does not have access to instantaneous emission intensitities but only to the integrated intensity emission spectra.
 It is, thus, rather difficult to directly control equilibration.
Notwithstanding it has been possible to obtain clear signatures of RSB across the  lasing threshold in experiments on a particular subset of random lasers, having a quenched scattering structure and displaying large shot-to-shot spectral fluctuations.

%%%%%%%%%%%%%%%%%%%
%%%%%%%%%%%%%%%%%%%
\section{Experimental measurements of the Parisi order parameter}

In this section we will describe and discuss the first  experimental demonstration of RSB reported in ref.~\citeonline{Gho15} and give an overview of the numerous interesting results that have been published after this  work.
 A subsection will be dedicated to material  requirements for reproducing real replicas in random lasers.

\subsection{Experimental procedure}
\label{SS-EXP}
In Ref.~\citeonline{Gho15} the authors measure shot to shot intensity of the emitted light from a suitable random laser, calculate the fluctuation overlap $\mathcal C$ of these experimental observables and build its distribution  at various input pumping. The so obtained $P(\mathcal C)$ profiles unambigously reproduce the  predictions of RSB theory made by Giorgio Parisi. 

\begin{figure}[t!]
\centerline{\includegraphics[width=12cm]{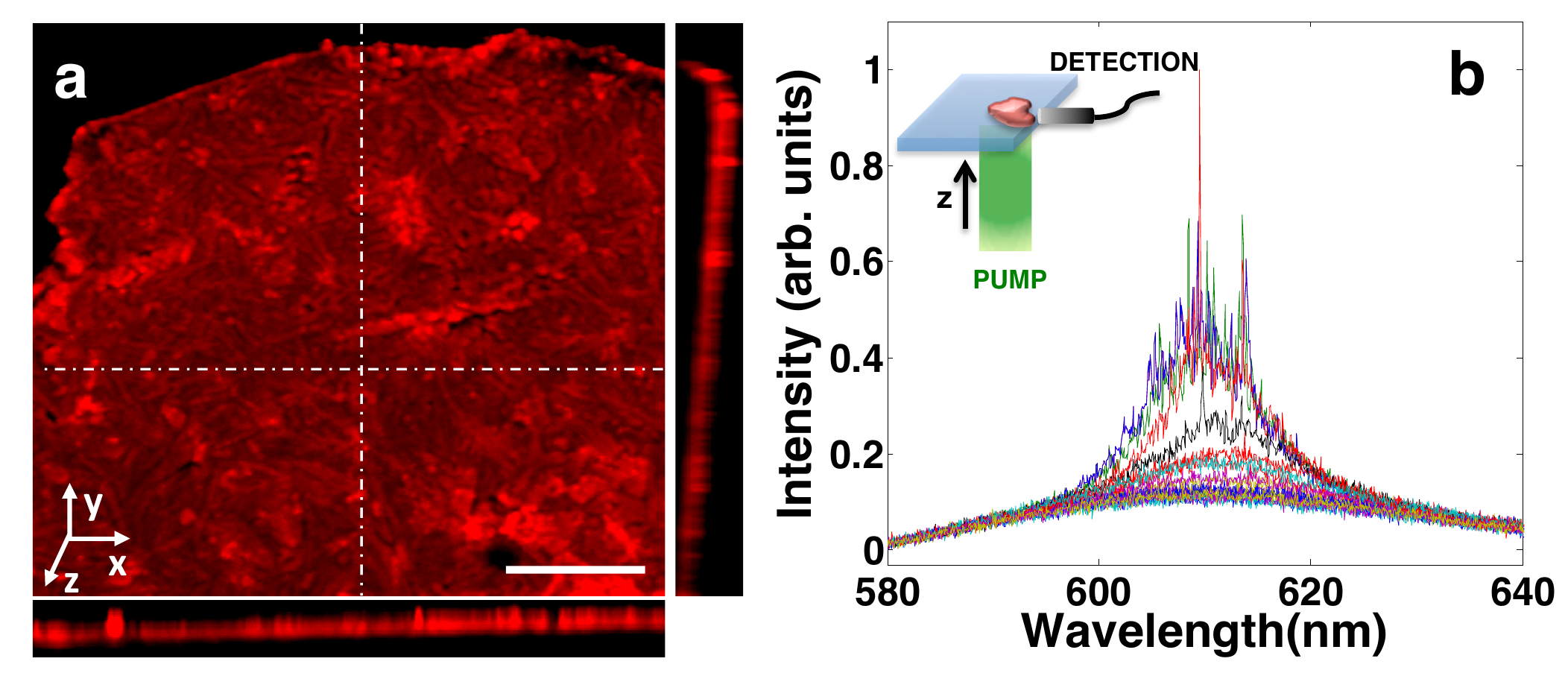}}
 \caption{
 a) 3D-reconstruction of confocal microscopy Z-stack images of  a T5OCx solid sample.
The right and the bottom panels report the yz- and the xz- sections, respectively. Scale bar: 20$\mu$m.
b) High resolution single shot spectra taken in the same conditions, 10mJ pump energy. Inset: sketch of the experiment.
Reproduced from Ref.~\citeonline{Gho15} with permission from  Nature Publishing Group.}
\label{fig-crystal}
\end{figure}

The authors use a thiophene (T5OCx) dye chemically treated  to have a thick amorphous solid material.
 When this material is pumped by an external source, above a certain threshold the fluorescence  is amplified and a stimulated emission  
 is obtained~\cite{Ghofraniha13,Anni03,Pisignano02}. 
 The amplification is sustained by the multiple light scattering inside the disordered system. 
A representative confocal image of the sample is depicted in fig~\ref{fig-crystal}a.
 For the investigation of emission fluctuations the sample is pumped by an external pulsed laser and at each pulse the emission  is collected.
A sketch of the pumping and collecting geometry is given in the inset of fig.~\ref{fig-crystal}b.

  %--------------Figure ------------------------%
%----------------end figure -----------------%

\begin{figure}[t!]
\centerline{\includegraphics[width=12cm]{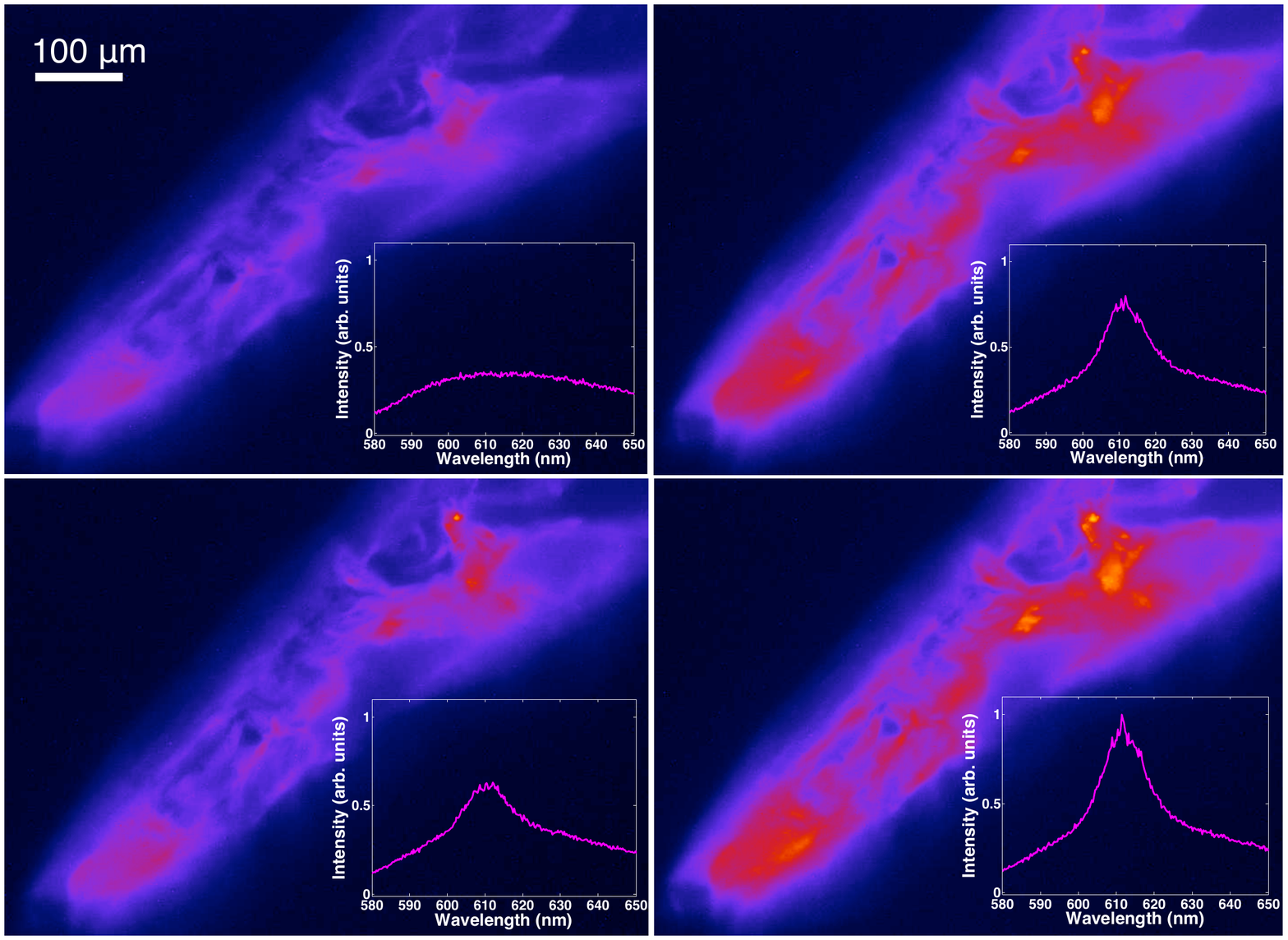}}
 \caption{
Snapshots of RL emissions.
Single shot optical images and corresponding emission spectra (insets) during the pumping of the sample in the same experimental conditions. The input energy is 10mJ.
Reproduced from Ref.~\citeonline{Gho15} with permission from  Nature Publishing Group. }
\label{fig-emissionimage}
\end{figure}

Fig.~\ref{fig-crystal}b shows single shot  RL emission spectra, taken at identical experimental conditions.
The distinct peaks are the activated modes or resonances of the disordered laser.
Their configuration  changes from shot to shot evidencing that
each time the system is pumped the numerous passive modes, characteristic of the material,  compete  for  the available gain,
giving rise to  several different compositions of the activated spectral peaks~\cite{Muj07}.
This is also evidenced by the direct visualisation of the sample during pumping, as  reported in fig.~\ref{fig-emissionimage}, where four different
fluorescence images
taken at four single shots  exhibit different emission patterns and spectra.
From shot to shot the size and the brightness of the luminous spots change showing that also the
spatial structure of the modes changes. Such behaviour is  strongly dependent on the input energy and it occurs above a pumping threshold that is the RL energy threshold. 
Below this pumping energy 
 only spontaneous emission (fluorescence) is observed, above this threshold the system becomes a RL with strongly variable emission. 

%--------------Figure ------------------------%
%----------------end figure -----------------%

\begin{figure}[t!]
\centerline{\includegraphics[width=12cm]{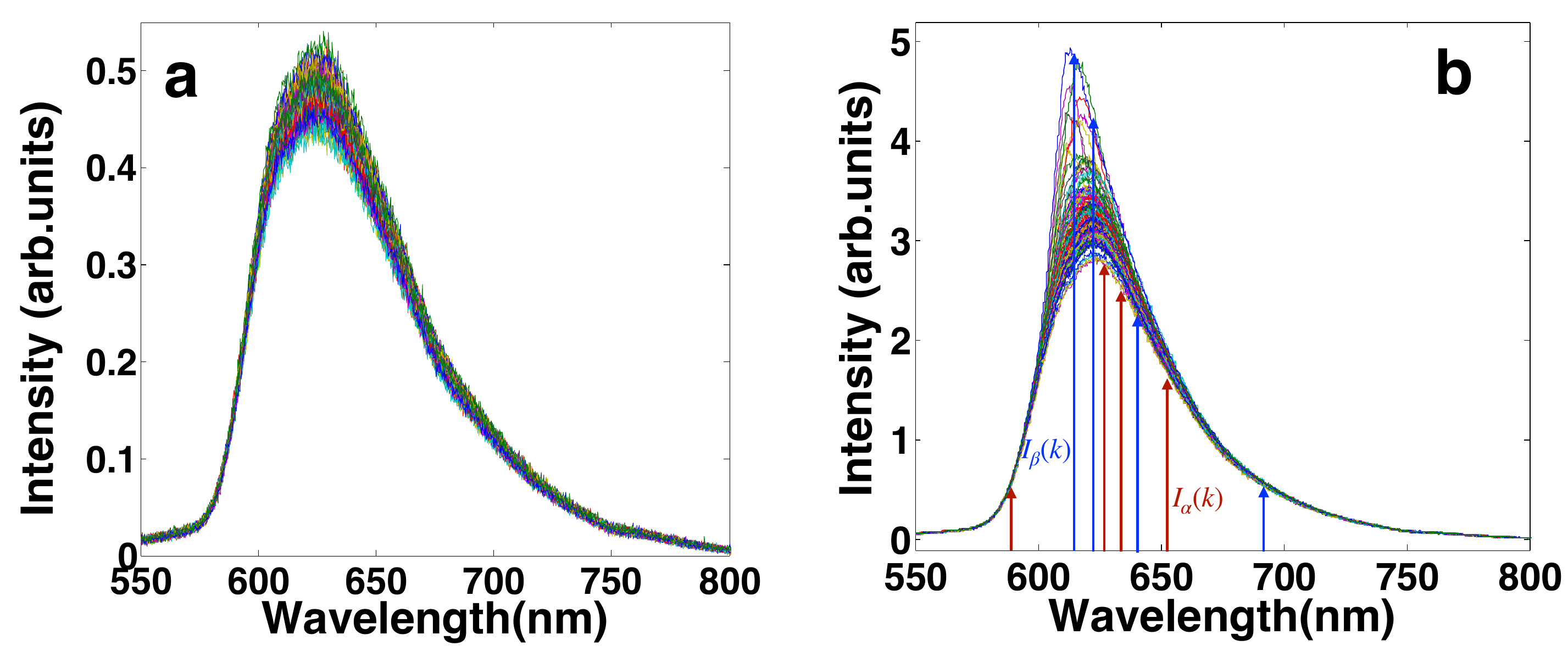}}
 \caption{
 a-b) Emission spectra at low energy 1mJ (a) and high energy 12mJ with evident fluctuations (b).
The experimentally accessible variable, coarse graining the behavior of single modes, is the intensity $I(k)$  at a given wavelength $\lambda_k$. In figure b the resonances $I(k)$ for two and for different replicas (spectra), denoted by  $\alpha$ (red arrows) and $\beta$ (blue arrows), are pointed out at.}
\label{fig-emission}
\end{figure}

In fig.~\ref{fig-emission}  emission spectra of subsequent 100 shots  at two different pump energies are shown.
At low energy (fig.~\ref{fig-emission}a) only  noisy variations of the spontaneous emission are observed, while 
at high energy (fig.~\ref{fig-emission}b) the spectra fluctuate randomly from pulse to pulse. 
These experimental results can be analyzed in the framework of replica  theory. Indeed, the solid property of the material guarantees the reproducibility of the same disordered network of mode interactions from shot to shot. Bond-disorder is quenched and real replicas of the same random laser sample are reproduced at each illumination shot. 
Each spectrum can, thus, be considered as the intensity configuration of a different state
of the same thermodynamic glassy phase.
Though for the theory, cf. Eqs. (\ref{d-leading-hamiltonian}), (\ref{eq-Ham-leading}), the spins are complex mode amplitudes, 
the experimental accessible observables are their intensities (\ref{d-intensity}), $I_k^{\rm a}$,
at wavelength $\lambda_k$ of the shot $\rm a$:
$k$ is the spin index and a is the replica index.

%--------------Figure ------------------------%
%----------------end figure -----------------%

It is important to recall that  the effective statistical
 mechanics Hamiltonian  variables are  the complex
 amplitudes $a_{k}$, \cite{Leu09,Con11,Ant15c} not  accessible in
  the experiments. While, as anticipated in section \ref{SS-NB}, their square moduli, the emitted intensities, are easily measurable. 
  In section \ref{SS-IFO}
we reported that  such coarse-graining is refined
 enough to validate the possible breaking of replica symmetry in random lasers and  their glassy-like behaviour by introducing the IFO parameter $\mathcal C$, Eq. (\ref{d-IFO}),   between pulse-to-pulse intensity fluctuations in different  replicas. Now we are considering experimental replicas and the experimental IFO is naturally  defined as 
\begin{equation}
\label{overlap}
\mathcal C_{\rm ab}= \frac{1}{\mathcal N_{\rm ab}}
\sum_{k=1}^N \Delta_k^{({\rm a})}  \Delta_k^{({\rm b)}},
\end{equation}
where 
\begin{equation}
\label{Delta}
 \Delta_k^{({\rm a})} \equiv I_{k}^{(\rm a)}-\bar I(k)
\end{equation}
with $\bar I(k)$ the average over $N_s$ replicas (emission spectra) of each  mode intensity
\begin{equation}
\bar I_k=\frac{1}{N_s}\sum_{\rm a=1}^{N_s}I_k^{(\rm a)}
\end{equation} 
and where the normalization factor is 
\begin{equation}
  \mathcal N_{\rm ab} \equiv  \sqrt{\sum_{k=1}^N  \left(\Delta_k^{({\rm a})}\right)^2  } \sqrt{\sum_{k=1}^N \left(\Delta_k^{({\rm b})}\right)^2 }
  \label{IFO-EXP-NORM}
\end{equation}

\begin{figure}[t!]
\centerline{\includegraphics[width=.99\linewidth]{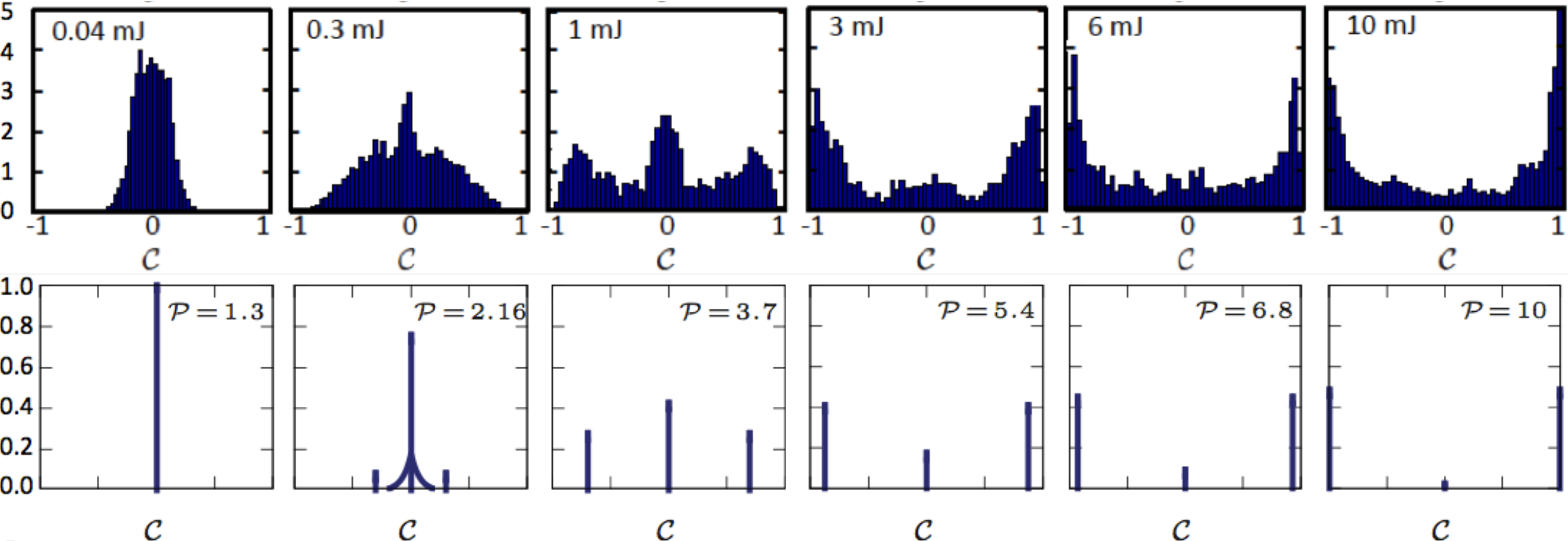}}
 \caption{
 Top line: distribution function of the experimental overlap $P(\mathcal C)$, Eq. (\ref{overlap}), showing replica symmetry breaking by increasing pump energy. 
 Bottom line: distribution of the theoretical IFO (\ref{d-IFO}) by increasing pumping.
 Reproduced from Ref.~\citeonline{Gho15} and \citeonline{Ant15a} with permission from  Nature Publishing Group.}
\label{fig-RSB}
\end{figure}

From the $N_s$  measured spectra  the set of all $N_s(N_s-1)/2$ values of $\mathcal C$   for  each different input  energy is calculated
and their distributions $P(\mathcal C)$ are depicted in fig.~\ref{fig-RSB} (top).

%--------------Figure ------------------------%
%----------------end figure -----------------%
%--------------Figure ------------------------%
\begin{figure}[t!]
\centerline{\includegraphics[width=.6\linewidth]{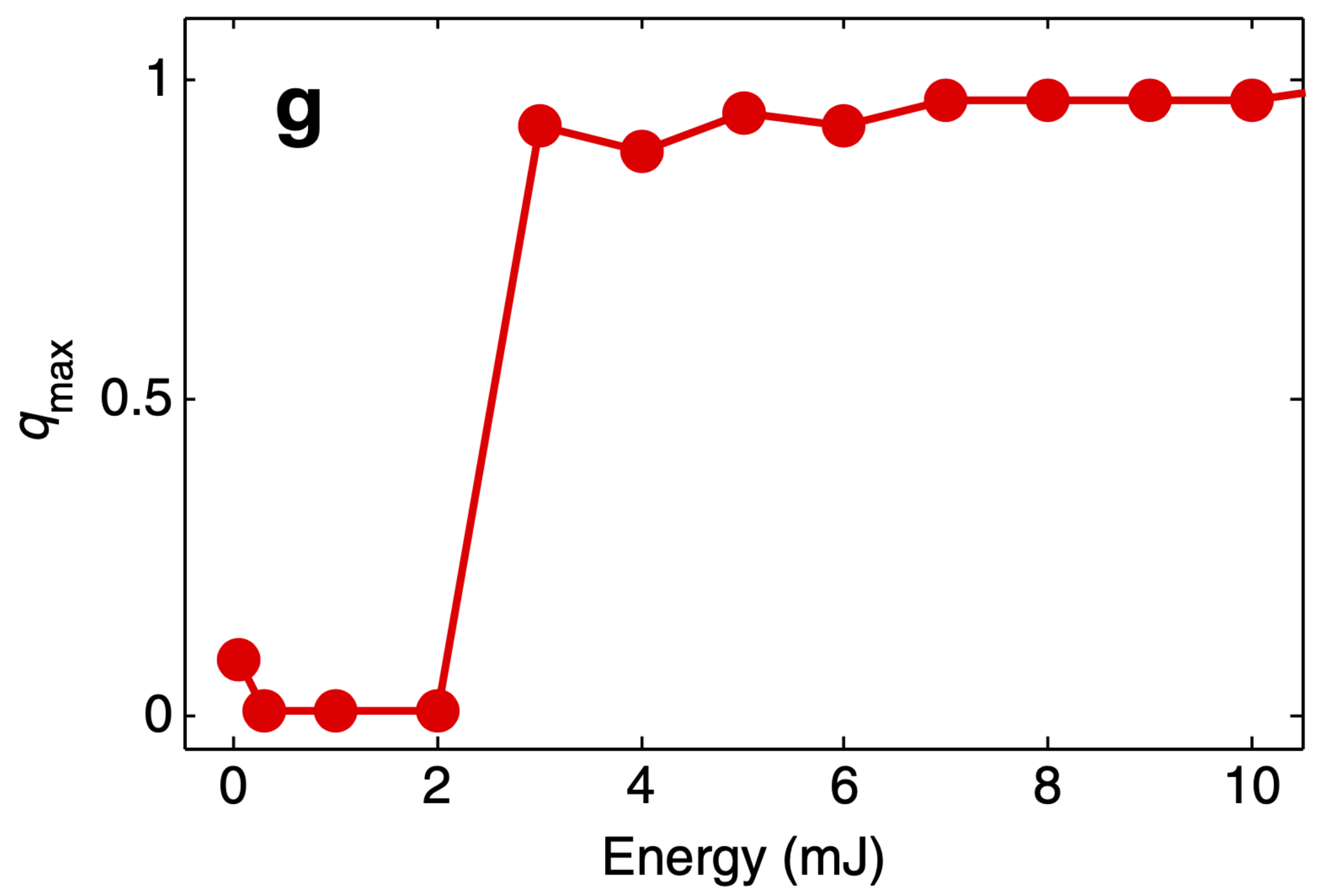}}
 \caption{
 The value $q_{\rm max}$ corresponding to the position of the maximum of $P(|\mathcal C|)$ versus pumping. A sharp transition from $0$ to a large value  is evidenced.
 Reproduced from Ref.~\citeonline{Gho15} with permission from  Nature Publishing Group.}
\label{fig-qmax}
\end{figure}
%----------------end figure -----------------%

Six examples are reported in fig.~\ref{fig-RSB}, top row of panels, for increasing pump
energy. At low energy (fig.~\ref{fig-RSB} left-most plot), all overlaps are centred
around the zero value, meaning that the electromagnetic modes (spins) are
independent and not interacting, they are in the paramagnetic regime.  By
increasing energy, modes are coupled by the nonlinearity and this
corresponds to a non-trivial overlap distribution.  In the high energy glassy phase, with all modes
highly interacting and frustrated by the disorder, $\mathcal C$ assumes all
possible values in the range $[-1,1]$.  Such behaviour of the $P(\mathcal C)$
evidences the fact that the correlation between intensity fluctuations
in any two replicas depends on the replicas selected. The variety of
possible correlations extends to the whole range of values. This is a
manifestation of the breaking of the replica symmetry. In the bottom row the analytical results of the fully connected model of sections \ref{SS-NB}-\ref{SS-IFO} are reproduced for a heuristic comparison.

In fig.~\ref{fig-qmax} 
a useful parameter $q_{\rm max}$ corresponding to the position of the maximum of
$P(|\mathcal C|)$ versus pumping is shown: it changes drastically from $0$ to a very large valure next to $1$ signaling a phase
transition at about  3mJ, this is the RL energy threshold. 
As shown in figures  (\ref{fig-phdi}) and (\ref{fig-Pq-PC-num}), the onset of the transition to the glassy phase occurs as soon as the $P(\mathcal C)$ develops non-gaussian tails. Taking as a reference the Full Width Half Maximum (FWHM) of the low pumping gaussian distribution one might, therefore, improve the identification of the random lasing threshold as the pumping energy at which 
the area outside the range of $2$ FWHM is sensitively larger than $2\%$.

These findings besides being a first experimental demonstration of the RSB theory, are also relevant in photonics.
The overlap distributions and $q_{\rm max}$ are now used in many works to measure the energy threshold of RLs and  are considered a powerful tool
to identify the RL behaviour  and distinguish it from other emission mechanisms.

\subsection{Material requirements for reproducing real replicas.}

Before reporting the main results of RSB in other kinds of RL, it is important to stress the concept that not all random lasers show strong emitted intensity fluctuations and consequent 
breaking of replica symmetry. 
In  ref.~\citeonline{Gho15}, for instance, it is demonstrated the lack of RSB  in a RL made of a dyed colloidal dispersion. 
Indeed, in a   fluid sample the particles tend to move from shot to shot during a single experiment and, therefore, the set of possible multiple scattering trajectories experienced by the light pumped into the system change. This implies that the optical susceptibility, as well as the normal mode profiles in (\ref{def-J2}) and (\ref{def-J4}) change giving rise to a different quenched disordered sample every shot. 
Under this conditions a RL sample is not replicated but a new RL, with a new random coupling  
configuration, is realized at every shot,
unless the liquid is not stable enough (from the point of view of photonic time-scales)  to allow the realization of replicas. 
Within this discussion, we report that RSB has been evidenced in liquid phase RL samples for instance in Ref.~\citeonline{Pin16} and Ref.~\citeonline{Tom16}.

In Ref.~\citeonline{Pin16}        
 a dyed sol-gel colloidal suspension with modified  amorphous TiO$_2$ scattering particles, made on purpose to strongly hinder photodegradation and precipitation, shows clear glassy random laser behaviour.
The spectra are  analysed by following the  procedure reported in section \ref{SS-EXP} and RSB is evidenced. 
Although the authors do not deepen on the fact that in liquid materials the structural composition changes in time due to the brownian motion of the particles,  they directly test  the robustness of their samples against tens of thousand of shots and  give evidence that emission spectra from single pump shots can be properly considered as replicas. As pumping increases from the incoherent wave regime across the lasing threshold the $P(\mathcal C)$ behaviour  is, indeed, qualitatively similar to the one reported in figure \ref{fig-RSB}: a low pumping  $P(\mathcal C)$ with a single gaussian peak in zero develops long tails around the lasing threshold that become side peaks at larger and larger IFO values upon increasing the pumping. 

The opposite approach, pumping energy on a fluid RL whose  microscopic scatterers position certainly changes from shot to shot, is followed in Ref.~\citeonline{Tom16}. 
Experimental evidence is provided  of the motion of the scattering particles from shot to shot 
 and, looking at the IFO distribution,   the robustness of RSB theory also in systems with
annealed disorder is claimed. 
If the statistical mechanical description of glassy random lasers as 
spin-glass models is to hold, however, one would expect the occurrence of a behavior corresponding to a replica symmetric solution \cite{Chen15}.
Our point of view is that in solid systems single shot spectra are considered as replicas because all experimental parameters are quenched and only the strong interaction between modes (spins) 
at high pumping (low temperature) causes the breaking of their symmetry.
In fluid materials whose typical diffusion time-scale is shorter than the experimental time (even though possibly longer then the single shot duration) the microscopic matter composition changes from shot to shot. Even though still  many modes are randomly activated and interact with quenched disordered couplings at any shot and the system is a random laser it is not possible to look for replica symmetry breaking because real replicas (i.e., same realization of the random coupling network) are not there.

Looking at the IFO distributions reported in Ref. \citeonline{Tom16}, indeed, 
thery turn out to have a sharp change from a gaussian to a purely bimodal distribution across the lasing threshold.  A more likely interpretation of this phenomenon might, thus, be the occurrence of bistability, a known phenomenon in standard lasers.
Because of randomness (at each shot a different realization of disordered mode couplings is yielded) some samples will be lasing after an illuminating shot, while some other might still be in the fluorescent regime. This implies that the average spectrum $\bar I_k$ over all emissions will never be similar to any of the single emission spectra: it will be far away from both the fluorescent spectra and the lasing emission spectra.
Therefore, no $\Delta_k$ in Eq. (\ref{Delta}) will be small and the only values of the IFO available will be large (positive or negative), inducing a bimodal $P(\mathcal C)$.

 This is also what probably occurs in some other experiments on ordered lasers.

\subsubsection{RSB in ordered cavity}
We mention two very bright 
 works where RSB has been claimed to occur in standard  ordered cavities.
 Basak et al. in 2016 report on strong intensity fluctuations in both liquid and solid dye lasers with Fabry-Perot cavities obtained by the cuvette walls in the former and 
 by the interface of polymeric thin slab with air in the latter~\cite{Bas16}. 
 The samples are pumped with pulsed  lasers and
 by increasing the pumping energy at laser threshold strong intensity fluctuations  are observed. These are analyzed by means of the  IFO distribution and 
RSB  is put forward as an explanation of their behaviour.

Actually, the very detailed analysis performed by the authors demostrates the existence of a critical interval of pumping energy where, from shot to shot, a fluorescent or a laser emission take place.
This is a clear indication of bistability, in laser language, corresponding to phase separation in a first order phase transition, in statistical mechanics. 
It is not unexpected, as the statistical physics model   for the ordered (or for the not too disordered) multi-mode laser  in a closed cavity corresponds to
Eq. (\ref{d-leading-hamiltonian}) with only the four-mode coupling part and with small enough $R_J$ for the $J$'s distribution, cf. figure \ref{fig-phdi}.

Another example is the work of  Moura et al. in Ref.~\citeonline{Mou17}  where
 the IFO distribution is measured in the spontaneous mode-locking
regime of a multimode Q-switched Nd:YAG laser. 
 The authors rightly assert that the observed phenomenon is quite distinct from
what investigated in random lasers characterized by incoherently oscillating modes. 
As they display  the $P(\mathcal C)$, calculated via (\ref{overlap}), (\ref{Delta}) and (\ref{IFO-EXP-NORM}), vs pumping  the lasing threshold is identified as the sharp transition from a phase of a single gaussian peaked $P(\mathcal C)$ to a clean bimodal $P(\mathcal C)$.

Since in Ref~\citeonline{Mou17} 
the ensemble of spectra from which the $P(\mathcal C)$ is computed is very instructively displayed, it is, therefore, possible to see that above threshold only two types of spectra are present: a broad, flat fluorescent one and a narrow, spiky lasing one. Once again, we face bistability.
This is expected in the theory transition between model-locking lasers (ordered or slightly random) as reported, for an instance, by the dotted line in figure \ref{fig-phdi}.

From this latter study we also report a interesting phenomenon. As pumping is large enough so that  all shots are lasing, strong fluctuations appear to be there in the (narrow) spectra and the $P(\mathcal C)$ appears to have a more complicated shape, though always basically bimodal, in which overlaps between the side peaks have a finite probability to occur.
It would be interesting to understand whether
this is due to experimental noise correlated with the strong pumping or it is the onset of a glassy phase as, for instance, in Fig. \ref{fig-phdi} for $0.5<R_J<1$.
 
\subsection{Evidence of RSB in other different RLs and nonlinear waves}

\begin{figure}[t!]
\centerline{\includegraphics[width=13cm]{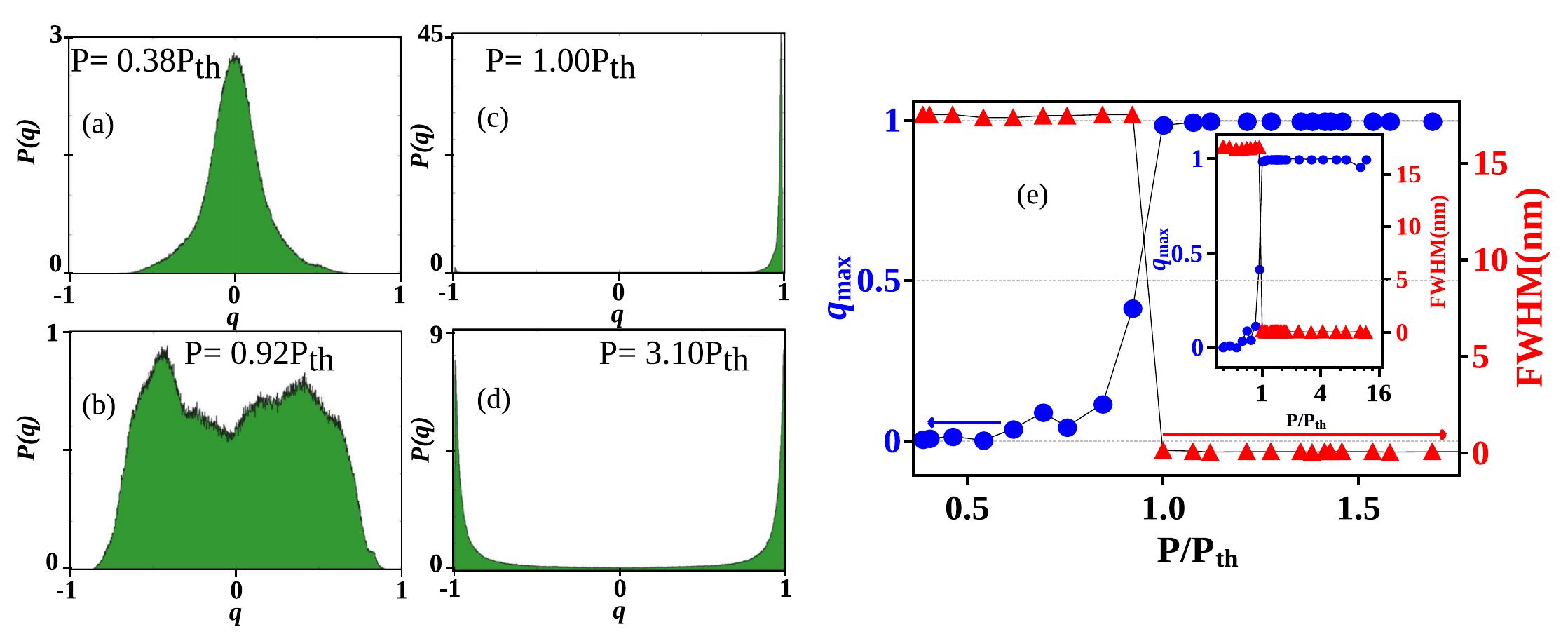}}
 \caption{ a-d) IFO distribution  $P(q)$ obtained from experimental data at different pump powers (normalized with respect to the threshold power $P_{\rm th}$). e) $q_{\rm max}$ (circles) versus  normalized pump power, together with the FWHM (triangles) for  comparison. The inset shows the results for pump powers up to 12$P_{\rm th}$. 
 Reproduced from Ref.~\citeonline{Gom16} with permission from  American Physical Society.}
\label{fig-GomesPRA_2}
\end{figure}

A completely different class of RLs showing RSB is furnished by erbium-based random fiber lasers (Er-RFL) and  widely investigated by A. Gomes and coworkers.
A first demonstration in  given in ref.~\citeonline{Gom16}, where the authors
employ  a 30-cm-long
Er-RFL, pumped by a continous wave source.
The authors discuss and show that this kind of RFL is a multimode laser with more than 200 longitudinal modes. They analyse 
 emission intensity fluctuation over 1500 acquisition spectra (replicas) for different pumping power and measure the IFO  from Eq.~(\ref{overlap}).
In fig.~\ref{fig-GomesPRA_2} the IFO distribution $P(q)$ is depicted together with the parameter $q_{\rm max}$, at which $P(q)$ assumes its global maximum. \footnote{We warn the reader that the  symbol $q$, rather than $\mathcal C$, is used here and in the following to denote the intensity fluctuation overlap.} 
It is demonstrated in fig.~\ref{fig-GomesPRA_2}-e that both sharp line narrowing and $q_{\rm max}$ growth from $0$ to $1$ 
occur at the same identical input power interval, that is the laser threshold. This confirms the findings in Ref.~\citeonline{Gho15}.

%--------------Figure ------------------------%
%----------------end figure -----------------%

\subsubsection{RSB and  L\'{e}vy flight}

Transition from paramagnetic to glassy states in random lasers has been also analyzed in correspondence  to the transition from diffusive to super diffusive (or L\'{e}vy flight) regimes of light propagation. 
Strong emission fluctuations in RLs have been described in the past by using the diffusive model of light and by successfully identifying  different statistical regimes for disordered lasers~\cite{Lep07,Lep13}. 
Recently, the strong emission fluctuations in RLs have been analyzed both in the frameworks of    L\'{e}vy statistics and spin glass theory.
Interesting similarities and differences have been demonstrated.

A first comparison between the two statistical physics properties  is reported by Gomes et al. in Ref.~\citeonline{Gom16a}.
In this work the physical origin of the possible
correspondence between the L\'{e}vy flight statistics of emission intensity  and the photonic RSB glassy transition
in RLs is both theoretically and experimentally investigated. 
The RL consists of crystalline powders of Nd$^{3+}$-doped YBO$_3$ (Nd:YBO) and pumped by a pulsed laser.
The authors for the first time explain the comparison between the RSB transition to the photonic glassy phase and the changes in the statistics
of intensity fluctuations in RLs within the same theoretical framework based on the Langevin equations describing the evolution of the mode amplitudes.
While such theoretical approach has been introduced and widely used by the spin-glass community~\cite{Angelani06a,Angelani06b, Leu09,Con11}, only recently it  has been considered for the statistical 
distribution of RL intensities~\cite{Rap15}. 
Starting from the  Langevin equation 
the probability density function of emission intensity is obtained. The steady state solution of such equation gives  
the L\'{e}vy-like distribution of intensities.
 The authors show that,  for a given disorder strength, by increasing pumping rate
 the statistics of emission intensities shifts progressively from an initial
Gaussian  to a L\'{e}vy-like  and, then, again to a  Gaussian 
regime, as predicted by the diffusive model of light propagation in random media.
The  IFO distribution  calculated  on the same experimental data show
 a transition from the spontaneous emission-paramagnetic to the RL glassy behaviour and a recurrence  to the paramagnetic regime at higher pumping, similarly to the L\'{e}vy  statistics results.
 Once again, this critical behavior is interpreted as a RSB transition, whereas,  looking both at the strict bimodal shape of the IFO distribution and at the series of spectra displayed in the work, the transition is probably a first order phase transition with coexistence of fluorescence and (random) laser phases. That is, bistability.
 
The novelty of these findings lies  at higher excitation pulse energy  well above the threshold, where fluctuations inside the single spectrum decline considerably 
with the consequent restoration of the Gaussian diffusive regime for the intensities, but the RL becomes more and more disordered from the point of view of mode coupling and the IFO distribution appears to become non-trivial, hinting the possible onset of a glassy random laser.
In this article the authors do not give any explanation of the existence of a strict causal link between the self-averaged Gaussian regime 
of intensity fluctuations above the threshold and the observed suppression of the glassy phase, leaving the subject to  further studies.
The link between the
onset of the  L\'{e}vy statistical regime of intensity fluctuations and
the emergence of the RL regime has been reported also in a   one dimensional Er-RFL~\cite{Lim17}.

A different result is shown in ref.~\citeonline{Tom16}, where the authors conclude that
the RL transition and the L\'{e}vy regime onset may not have a clear causal relation. 
They claim that the former is  strongly related to the threshold and less on the magnitude of the fluctuations, while the latter  emerges
for large fluctuations and  only under strict conditions of the ratio between gain and scattering properties of the material.
As a demonstration they report the clear experimental result  where
the  L\'{e}vy regime is suppressed while a RSB transition is present.
A further study on the matter is probably required, also including the role of bistability in non-glassy random lasers.

\subsubsection{RSB and  Turbulence}
In 2018 Gonzales et al. employ an erbium random fiber laser  to demonstrate for the first time
  the coexistence of turbulence-like and spin-glass-like behavior from the same set of measurements~\cite{Gon18}. 
 
For the  theoretical analysis they introduce the photonic Pearson correlation coefficient that can be considered as a generalized time dependent expression of 
 the spin-glass  overlap parameter. If  $\tau$ is the time of data acquisition, the time dependent IFO between replicas $\alpha$ and $\beta$ is defined as 
 \begin{equation}\label{Pearson} 
 Q_{\alpha\beta}(\tau)=\frac{\sum_k \Delta_k^{(\alpha)}(\tau) \Delta_k^{(\beta)}(\tau)} {\sqrt{\sum_k \left(\Delta_k^{(\alpha)}(\tau)\right)^2}\sqrt{\sum_k
 \left(\Delta_k^{(\beta)}(\tau)\right)^2}},
 \end{equation}
  with $k$ being the wavelength index in the emission spectra, $\alpha$ and $\beta$ single shot spectra  indexes (replicas).
Besides the definition (\ref{Delta}), for $\tau=0$, i.e.,  $$\Delta_k^{(\alpha)}(0)\equiv I_k^{(\rm a)}(0)-\bar I_k(0), $$ when  $\tau>0$ the time-dependent spectral fluctuation is defined as 

$$  \Delta_k^{(\alpha)}(\tau) \equiv I_k^{(\alpha)}(\tau)-I_k^{(\alpha)}(0) - \left[
\bar I_k(\tau)-\bar I_k(0)\right].
$$

  With these notations Eq.(\ref{Pearson}) quantifies the emission intensity fluctuations in time  and it is  sensitive to both liquid-glass transition and fluid dynamics phenomena such as  turbulence. 
It is demonstrated that above threshold  the distribution of the  Pearson coefficient for short times  is centered around $Q \sim 0$, 
leading to the unimodal behaviour of $P(Q)$ in the turbulent-like state. 
 While always above threshold but
at  large time scales it
assumes a distribution resembling the  profile typical of a RSB IFO distributions. 
In this regime
the intermittency vanishes and a crossover  to the non-turbulent behavior takes place.  

The authors  state that, since the overlap parameter considers all separation times between spectra, the statistical weight of the replica overlaps with 
long times dominates over the short time series. As a consequence, $P(Q(\tau))$ actually appears qualitatively similar to the overlap distribution that characterizes the RSB spin glass phase.

 \subsubsection{RSB maps}
\begin{figure}[t!]
\centerline{\includegraphics[width=.95\textwidth]{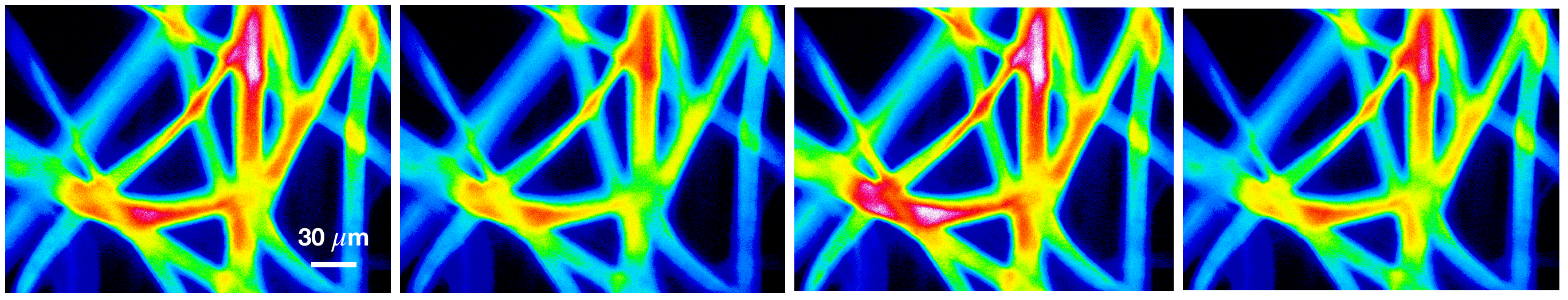}}
 \caption{
Four fluorescence images taken in identical experimental situations and regarded as replicas. 
Reproduced from Ref.~\citeonline{Mas21} with permission from  ACS Publications.}
\label{fig-images}
\end{figure}

Very recently RSB theory has been applied to obtain real time maps of the laser activity in a heterogeneous RL~\cite{Mas21}.
The random lasers are made of polymeric ribbon-like and highly porous fibers with evident RL action from separated micrometric domains that alternatively switch on and off by tuning the pumping light intensity. This novel effect is visualised by building for the first time replica symmetry breaking maps of the emitting fibers with micrometric spatial resolution.
The overlap parameter is calculated directly from the images. 100 single shot frames are recorded at fixed experimental condition and each one is a replica. 
Four examples are illustrated in fig.~\ref{fig-images}, where the variation of the spatial emission from shot to shot is evident.
In this work the measurable quantities corresponding to spins are the intensities of the pixels recorded by the camera. 
 The authors  calculate the  overlap between these observables that are the spatial (transverse) modes while previously only temporal (longitudinal) modes were considered~\cite{Gho15}.

 %--------------Figure ------------------------%
%----------------end figure -----------------%

\begin{figure}[t!]
\centerline{\includegraphics[width=.9\textwidth]{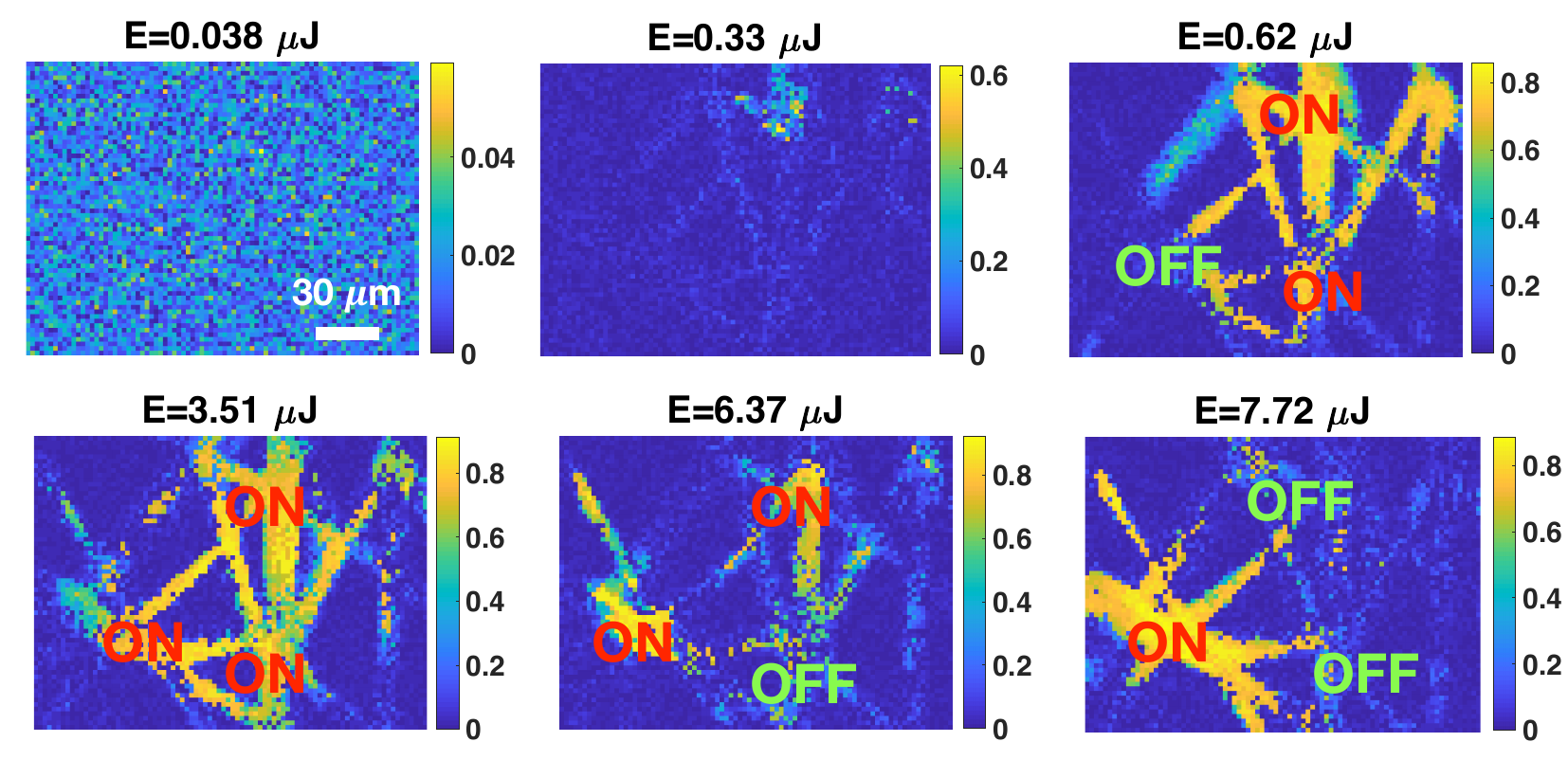}}
 \caption{
Replica symmetry breaking maps. Maps of the $q_{\rm max}$values calculated for each macropixel at different input energies. ON and OFF indicate the activation status of the RL emission.}
\label{fig-RSBmap}
\end{figure}

Each frame of 1024$\times$1376 pixels is divided into squares of 16$\times$16, named macropixels, and the overlap $q^j_{\alpha,\beta}$ for each macropixel $j$ is calculated as:
\begin{equation}\label{overlap-map}
q^{(j)}_{\alpha\beta}= \frac{\sum_{x_k,y_k} \Delta^{(\alpha,j)}(x_k, y_k) \ \Delta^{(\beta,j)} (x_k, y_k)}{\sqrt{\sum_{x_k,y_k}\left(\Delta^{(\alpha,j)} (x_k, y_k)\right)^2 } \sqrt{\sum_{x_k,y_k} \left(\Delta^{(\beta,j)} (x_k, y_k)\right)^2 }}.
\end{equation}
where $\alpha$ and $\beta$ are replica indexes, $j$ is the macropixel index, and $x_k$ and $y_k$ are pixel indexes inside one macropixel running from 1 to 16.
In eq.~(\ref{overlap-map})
$$
 \Delta^{(\alpha,j)} (x_k, y_k)\equiv I^{(\alpha,j)}(x_k, y_k)-{\overline {I^{(j)}}} (x_k, y_k),
$$
with $I^{(\alpha,j)}(x_k, y_k)$ being  the intensity at the pixel with coordinate $(x_k, y_k)$ of the $j^{\rm th}$ macropixel for the  replica $\alpha$ and where
$$
{\overline {I^{(j)}}} (x_k, y_k)\equiv \frac{1}{N_s}\sum_{\alpha=1}^{N_s}I^{(\alpha,j)}(x_k, y_k)
$$
is the average intensity over $N_s$ shots for each pixel.
With this procedure, from the images one has access to the sets of all $N_s(N_s-1)/2$ values $q^{(j)}$ of $q^{(j)}_{\alpha \beta}$ 
and they determine the distributions $P(q^{(j)})$ for all 64$\times$86 macropixels $j$. This is done for different  pumping energy. 
Following the previous work, the authors calculate $q_{\rm max}$
 for all macropixels  and
obtain replica symmetry breaking maps  at various pumping.  The results are shown in fig.~\ref{fig-RSBmap}. 
Domains with $q_{\rm max} > 0.5$ are laser-ON and those with $q_{\rm max} < 0.5$ are laser-OFF.

 %--------------Figure ------------------------%
%----------------end figure -----------------%

In this work the mapping of $q_{\rm max}$ allows for the first time the visualisation of heterogeneous RL with switching and variable activity. This procedure is proposed as a robust tool to identify the presence and the spatial extension of RL activity from fluorescence images.

 \subsubsection{RSB in nonlinear waves}
 
\begin{figure}[t!]
\centerline{\includegraphics[width=.99\textwidth]{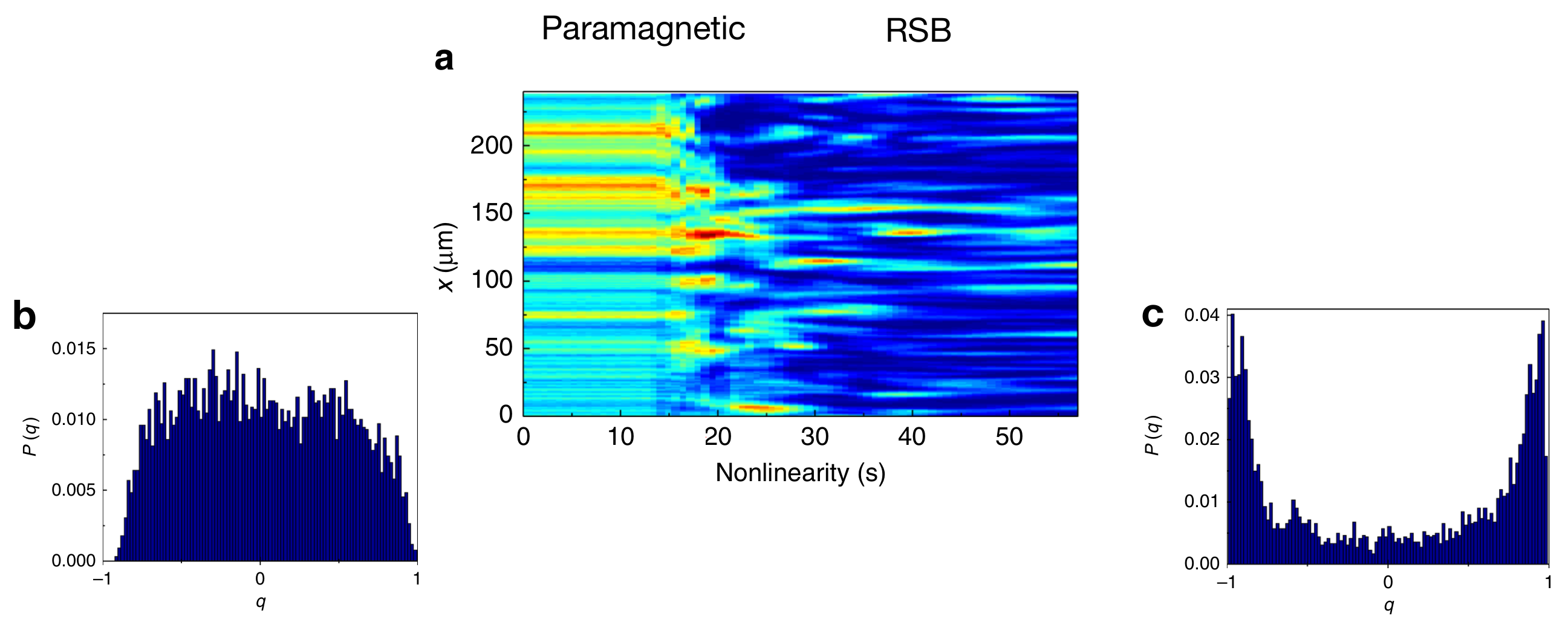}}
 \caption{
Transmitted intensity profiles along one direction $I(x)$ by increasing nonlinearity. At high nonlinearity RSB regime of strong fluctuations is evidenced. 
The corresponding distribution $P(q)$ in the paramagnetic (left) and glassy (right) regimes are depicted.}
\label{fig-Pierangeli}
\end{figure}

In 2017 Pierangeli et al.~\cite{Pie17} report the observation of the breaking of replica symmetry in nonlinear optical propagation. They investigate intensity profiles of a laser beam transmitted
from a photorefractive  disordered slab waveguide. The nonlinearity of the material is tuned and  strong  fluctuations of  light above a certain threshold are evidenced, as in the intensity map
 of fig.~\ref{fig-Pierangeli}.
 Differently from previous   works, the  replica overlap  is calculated over the spatial autocorrelation of the intensities $I(x)$,  and its distribution  undergoes a clear transition into a non-trivial distribution as the nonlinearity exceeds a threshold value. Results are shown in fig.~\ref{fig-Pierangeli}. These findings demonstrate that nonlinear propagation can manifest features typical of spin-glasses thanks to the coexistence and interplay of the  two main ingredients: disorder and nonlinearity.
 %--------------Figure ------------------------%
%----------------end figure -----------------%

%%%%%%%%%%%%%%%%%%%%%%%%%%%%%%%%
\section{Conclusions and Outlook}

In this chapter we have considered the problem of the experimental measure of the functional order parameter for a multiequilibria phase in complex disordered systems, representing the transition to a replica symmetry broken thermodynamic phase.
After a flash introduction of the Parisi overlap distribution we briefly discussed how the outcomes of the RSB theory have soon cast light on peculiar results in early experiments on spin-glass susceptibility. We have, then, been very rapidly reviewing the attempt to reconstruct the equilibrium overlap  distribution through measurements of the off-equilibrium fluctuation-dissipation ratio in systems satisfying stochastic stability and we did our best to give a state-of-the-art report of the experiments tried so far along that approach.  Finally, we have diffusively discussed, both theoretically, numerically and experimentally, the approach to the problem by means of complex random photonic systems, the random lasers, and reported the first measurements of the Parisi order parameter through the measurements of light intensity spectral fluctuations and the acquisition of overlaps between fluctuations of different real replicas of the systems.

The study of nonlinear photonics systems by means of statistical mechanics of disordered systems is an inspiring example of a constructive sinergy leading to advances in both fields.
The idea that random lasers might be described as disordered systems and treated with the tools of statistical physics in order to fully unveil and understand their complex behavior \cite{Angelani06a,Angelani06b,Leu09,Con11} has brought to the development of a spin-glass model\cite{Ant15b,Ant15c} whose RSB order parameter could finally be accessible to experimental measurements.\cite{Ant15a,Gho15,Gom16,Gom16a,Gomes21} 

Furthermore, such sinergy has stimulated several new studies on different open issues in photonics in random media. To mention a few, the statistical physics approach has been applied in relation to  the onset of a power-law (L\'evy-like) distribution of the emission intensity in random laser,\cite{Lep07,Lep13,Rap15,Gom16a,Tom16,Lim17} to the characterization of turbulence in photonic systems,\cite{Gon18} to the spatial distribution of interacting modes \cite{Antenucci21,Mas21} and the self-substained nature of mode-locking in random lasers \cite{Con11,Antenucci21} and has lead to the study of glassiness in nonlinear photonic systems other than random lasers \cite{Pie17}.
Several rather interesting issues remain open in this line of research combining photonics and statistical physics, including the study of  a first order critical behavior with phase coexistence and its relationship to the possible occurrence of bistability in random lasers, or the measurement and control of mode phases whose impact would have important consequences in the making and the technological use of random lasers.

%%%%%%%%%%%%%%%%%%%%%%%%%%%%%%%%%%%%%%%%
\section*{Acknowledgements}
The authors thank Fabrizio Antenucci, Miguel Iba\~nez-Berganza, Giacomo Gradenigo, Claudio Maggi, Alessia Marruzzo, Jacopo Niedda and Giorgio Parisi for very many useful scientific exchanges.

We acknowledge the support from the European Research Council (ERC) under the European Union’s Horizon 2020 Research and Innovation Program, Project LoTGlasSy (Grant Agreement No. 694925), the
support of LazioInnova - Regione Lazio under the program {\em Gruppi di ricerca 2020} - POR FESR Lazio 2014-2020, Project NanoProbe (Application code A0375-2020-36761).

\bibliographystyle{ws-rv-van}
\bibliography{RLexperimental-RSB40}

\printindex                         % to print subject index

\end{document}